\documentclass[aps,preprintnumbers,superscriptaddress,longbibliography,amsmath,amssymb,twocolumn,10pt,floatfix]{revtex4-1}

\usepackage{graphicx}
\usepackage{dcolumn}
\usepackage{bm}
\usepackage{xspace}
\usepackage{multirow}
\usepackage{booktabs}
\usepackage{xcolor}
\usepackage{xfrac}
\usepackage{makecell}

\usepackage{comment}
\usepackage{physics}
\usepackage[separate-uncertainty = true]{siunitx}
\DeclareSIUnit{\depth}{\gram\per\square\centi\meter}

\usepackage{hyperref}
\hypersetup{
    colorlinks=true,
    linkcolor=blue,
    filecolor=blue,
    urlcolor=blue,
    pdftitle={Overleaf Example},
    pdfpagemode=FullScreen,
    }

\usepackage[mathlines]{lineno}

\newcommand{\alphahad}{\alpha_{\text{had}}}
\newcommand{\alphaem}{\alpha_{\text{EM}}}
\newcommand{\zetahad}{\zeta_{\text{had}}}
\newcommand{\zetaem}{\zeta_{\text{EM}}}
\newcommand{\multhad}{m_{\text{had}}}
\newcommand{\multem}{m_{\text{EM}}}
\newcommand{\multtotal}{m_{\text{tot}}}
\newcommand{\elasticity}{\kappa_{\text{el}}}
\newcommand{\etahad}{\eta_{\text{had}}}
\newcommand{\etaem}{\eta_{\text{EM}}}

\newcommand{\conex}{\textsc{Conex}}
\newcommand{\epos}{E\textsc{pos\,}LHC\xspace}
\newcommand{\qII}{QGS\textsc{jet}\,-II.04\xspace}
\newcommand{\sibd}{S\textsc{ibyll}2.3d\xspace}
\newcommand{\eposr}{E\textsc{pos\,}LHC-R\xspace}
\newcommand{\qIII}{QGS\textsc{jet}\,-III.01\xspace}
\newcommand{\sibe}{S\textsc{ibyll}2.3e\xspace}

\newcommand{\xmax}{X_{\max}}
\newcommand{\dxmax}{\Delta X_{\max}}

\newcommand{\x}[1]{%
  {}$
  \kern-2\mathsurround 
  $
  \binoppenalty10000 \relpenalty10000 #1
  {}$
  \kern-2\mathsurround 
  $
}

\begin{document}





\title{Probabilistic mapping between multiparticle production variables and the depth of maximum in proton-induced extensive air showers}

\author{Lorenzo Cazon}
\address{Instituto Galego de Física de Altas Enerxías (IGFAE),\\ Rúa de Xoaquín Díaz de Rábago, s/n, Campus Vida, Universidade de Santiago de Compostela, 15705, Santiago de Compostela, Galicia, Spain}

\author{Ruben Concei\c{c}\~{a}o}
\address{Departamento de F\'isica, Instituto Superior T\'ecnico (IST), Universidade de Lisboa, Av.\ Rovisco Pais 1, 1049-001 Lisbon, Portugal}
\address{Laborat\'{o}rio de Instrumenta\c{c}\~{a}o e F\'{i}sica Experimental de Part\'{i}culas (LIP) - Lisbon, Av.\ Prof.\ Gama Pinto, 2, 1649-003 Lisbon, Portugal}

\author{Miguel A. Martins}
\email{miguelalexandre.jesusdasilva@usc.es}
\address{Instituto Galego de Física de Altas Enerxías (IGFAE),\\ Rúa de Xoaquín Díaz de Rábago, s/n, Campus Vida, Universidade de Santiago de Compostela, 15705, Santiago de Compostela, Galicia, Spain}

\author{Felix Riehn}
\address{Technische Universität Dortmund, August-Schmidt-Straße 4, 44221 Dortmund, Germany}

\date{\today}

\begin{abstract}
The interaction of ultra-high-energy cosmic rays with air nuclei triggers extensive air showers that reach their maximal energy deposition at the atmospheric depth $\xmax$. The distribution of this shower observable encodes information about the proton-air cross-section via fluctuations of the primary interaction point, $X_1$, and hadron production through $\dxmax \equiv \xmax - X_1$.
\par
We introduce new multiparticle production variables, $\alphahad$, $\zetahad$, and $\zetaem$, built from the energy spectra of secondaries in the primary interaction. Their linear combination, $\xi$, predicts over $50 \%$ of the fluctuations in $\dxmax$. Moreover, we build a probabilistic mapping based on the causal connection between $\xi$ and $\dxmax$ that enables model-independent predictions of $\xmax$ moments with biases below \SI{3}{\depth}. Therefore, measurements of the distribution of $\xmax$ allow a data-driven probing of secondary hadron spectra from the cosmic-ray-air interaction, in proton-induced showers.
\par
The distributions of the new multiparticle production variables can be measured in rapidity regions accessible to current accelerators and are strongly dependent on the hadronic interaction model in the kinematic regions exclusive to ultra-high-energy cosmic rays.
\end{abstract}

\pacs{Valid PACS appear here}
\maketitle


\section{Introduction} \label{sec:intro}
 Ultra-high-energy cosmic rays (UHECRs) with energies above $10^{18}\,$eV initiate extensive air showers (EAS) through interactions with atmospheric nuclei, producing secondary particles in kinematic regimes inaccessible to human-made collider experiments. The accurate description of these interactions is essential to infer the primary mass composition, and ultimately constrain the potential sources of UHECRs. However, the mass interpretation of EAS data is hampered by the lack of accelerator data in the relevant phase space~\cite{2012_Kampert_MassModelDep}, and is not consistently described by current phenomenological hadronic interaction models~\cite{2024_Auger_Xmaxs1000fits, 2022_Albrecht_MuonPuzzle}.
\par
The primary mass composition is typically inferred from the distribution of the depth of shower maxima, $\xmax$~\cite{2014_Auger_xmax}. Furthermore, the distribution of this shower observable is shaped by fluctuations of the first interaction point, $X_1$, and the stochastic production of hadrons in the primary interaction. The former fluctuations enable the measurement of the proton-air cross-section, $\sigma_{p-\text{air}}$~\cite{2012_Auger_xsection}, and the latter are captured by the distribution of $\dxmax \equiv \xmax - X_1$.
\par
Despite preliminary attempts~\cite{2023_Goos_HadInt_xmaxnmu, 2024_Martins_lambdamu_xmax}, there lacks an explicit connection between fluctuations of $\dxmax$ and hadron production in the primary interaction. Such correspondence independent of the hadronic interaction model, would enable data-driven constraints on the highest-energy hadronic interactions.
\par
In this work, we introduce new multiparticle production variables—$\alphahad$, $\zetahad$, and $\zetaem$—that capture fluctuations in energy partitioning among secondary particles of the cosmic-ray-air interaction, in proton-induced showers. The linear combination of these variables, $\xi$, predicts most of the fluctuations of $\dxmax$, providing a direct, model-independent probabilistic mapping between primary interaction physics and the distribution of $\xmax$. Furthermore, we analyse the kinematic properties of $\alphahad$, $\zetahad$, and $\zetaem$ and show that their distributions can be constrained by existing accelerator experiments in the far-forward rapidity region. These variables are highly sensitive to the shape of the energy spectra of hadrons in the phase space relevant for EAS development. Therefore, we show that $\xmax$ measurements can be used to constrain hadronic interactions beyond the reach of human-made colliders.
\par
The structure of this paper is as follows: In Section~\ref{sec:xmax_model}, we establish the connection between fluctuations in the primary-air interaction and $X_{\text{max}}$, for proton-induced showers. We derive a set of new multiparticle production variables and introduce their linear combination, $\xi$, as an estimator of $\dxmax$. In Section~\ref{sec:xi_model_performance} we evaluate the causal connection between $\xi$ and $\dxmax$ using detailed air shower simulations. In Section~\ref{sec:univerality}, we discuss the universality of the shower response to $\xi$ and propose a complete probabilistic model for the distribution of $\xmax$. In Section~\ref{sec:new_variables}, we interpret the new variables in terms of hadronic interaction physics, and in Section~\ref{sec:new_variables_accelerators} we explore their measurement in accelerator experiments. Our conclusions are presented in Section~\ref{sec:conclusions}.

\section{Connecting fluctuations in the primary interaction to fluctuations of \boldmath{$X_{\max}$}} \label{sec:xmax_model}

\subsection{The Heitler-Matthews framework} \label{subsec:review_hm_model}
The Heitler model~\cite{1936_Heitler_Hmodel} is a simplified description of photon-induced electromagnetic (EM) cascades as a branching process driven by pair production and \textit{bremsstrahlung}. Within this framework, the mean depth of the shower maximum, $X_{\max}^\gamma$, depends on the energy of the primary photon $E_0$, as
\begin{equation} \label{eq:heitler_xmax}
    X_{\max}^\gamma =  \lambda_r \ln \left( \frac{E_0}{\xi_c^e} \right),
\end{equation}
\par\noindent
where $\lambda_r \simeq \SI{37}{\gram\per\square\centi\meter}$ is the radiation length~\cite{2016_Gaisser_CRbook} and $\xi_c^e \simeq \SI{88}{\mega\electronvolt}$~\cite{2016_Gaisser_CRbook} denotes the electron critical energy, below which ionization losses dominate over radiative losses. Moreover, the model predicts that the number of particles at the shower maximum scales linearly with the primary energy $N_{\max} = E_0 / \xi_c^e$. Among other aspects, the model neglects the stochastic fluctuations in the showering process.
\par
The dynamics of pionic cascades can be understood by extending the Heitler model to hadronic showers. Within this Heitler-Matthews framework~\cite{2005_Matthews_HMmodel}, a primary proton with energy $E_0$ triggers a shower that reaches its maximum at
\begin{equation} \label{eq:hm_xmax}
    X_{\max}^p = X_1 +  \lambda_r \ln \left ( \frac{E_0}{2 \multtotal \xi_c^e} \right),
\end{equation}
\par\noindent
where $X_1$ denotes the depth traversed by the incident proton before interacting and $\multtotal$ denotes the total number of secondary particles, which is assumed to be constant throughout the cascade, irrespective of the nature and energy of the hadronically interacting particle. Moreover, the energy of each incident particle is assumed to be equi-partitioned between secondaries, in all interactions. The interaction depth and the proportion of charged to neutral pions are also assumed to be fixed. Lastly, Equation~\eqref{eq:hm_xmax} only considers the decay of neutral pions produced in the primary interaction.
\par
Despite this multitude of assumptions, Equation~\eqref{eq:hm_xmax} correctly describes the evolution of $\expval{X_{\max}^p}$ with the primary energy~\cite{2005_Matthews_HMmodel}, the so-called \textit{elongation rate}.

\par
Crucially, the Heitler-Matthews model does not account for the stochastic shower-to-shower fluctuations of $\xmax$, although they contain useful information about the multiparticle production of the primary interaction~\cite{2024_Martins_lambdamu_xmax, 2023_Goos_HadInt_xmaxnmu}.

\subsection{Derivation of a new multiparticle production variable of the primary interaction} \label{subsec:xi_derivation}
\begin{figure}[h!]
    \centering
    \includegraphics[width=\linewidth]{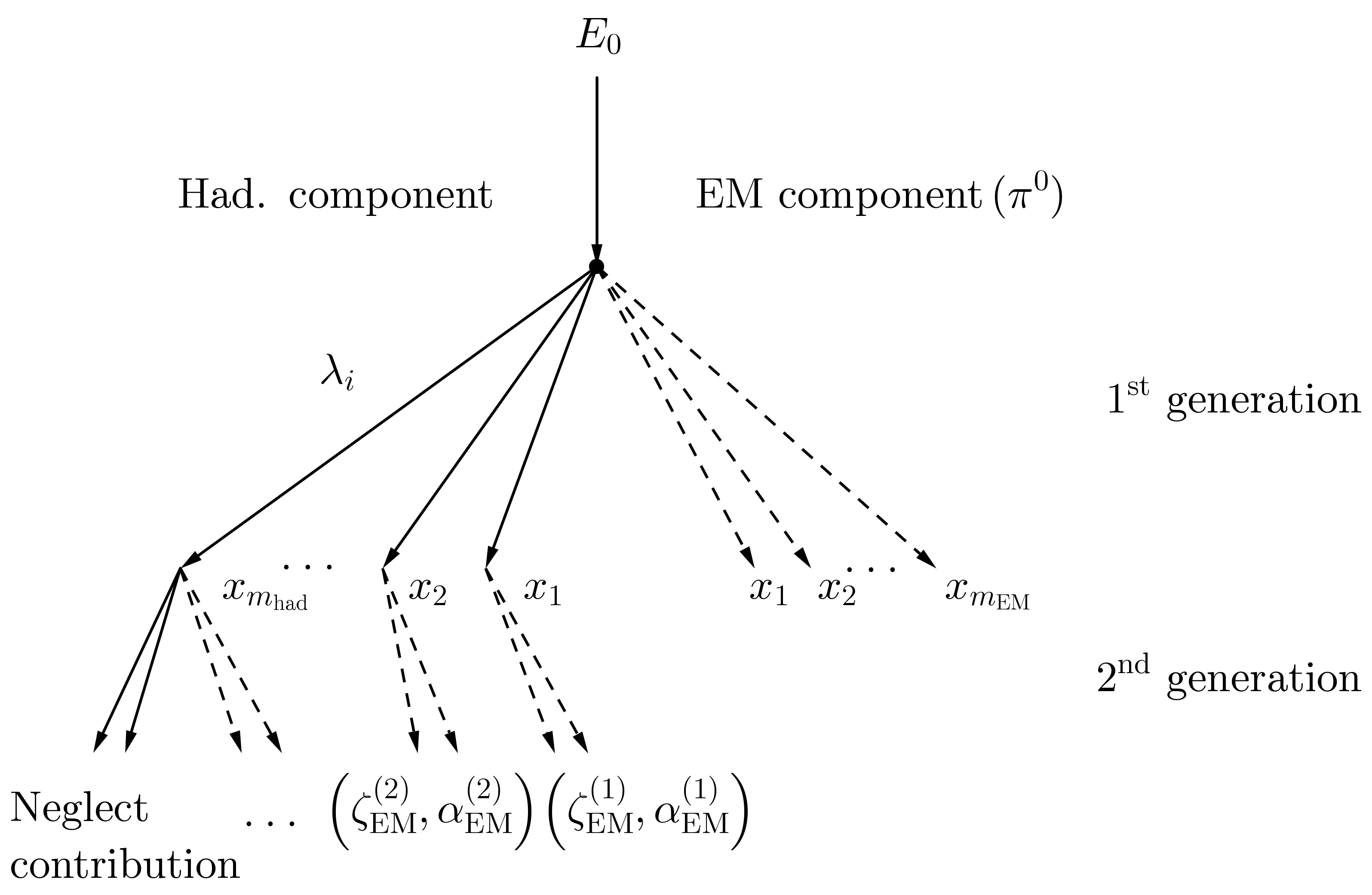}
    \caption{Scheme of the primary interaction, guiding the deduction of the model for fluctuations of $\Delta X_{\max}$. The variables $x$ represent the fraction of the primary energy carried by each secondary, in the laboratory frame. The number of hadronically interacting particles is given by $\multhad$, and that of neutral pions is $\multem$. Each hadronically interacting particle further interacts, after a depth $\lambda_i$, producing a subsequent EM cascade characterized by the multiparticle production variables $\alphaem$ and $\zetaem$, defined in the main text.}
    \label{fig:xi_model_scheme}
\end{figure}

We account for the shower-to-shower fluctuations of $\dxmax$ due to the stochasticity of particle production in the primary interaction by applying the Heitler-Matthews formalism to each secondary particle. Since our derivation is suited to proton-induced showers, the primary-air interaction will be referred to as primary or first proton-air interaction. A similar reasoning applied to the fluctuations of the number of muons, $N_\mu$, was explored in~\cite{2018_Cazon_alpha, 2023_Goos_PhDthesis}. We treat separately the EM cascades initiated by the decay of neutral pions produced in the primary interaction (first-generation neutral pions) and the EM cascades triggered by neutral pions produced when hadronically interacting secondaries of the primary interaction interact again (second-generation neutral pions). This is schematically represented in Figure~\ref{fig:xi_model_scheme}.
\par
\par
We start by examining the electromagnetic cascade induced directly by secondaries of the first $p$-air interaction. For this component, we only consider the contribution of neutral pions, as they account for $90 \%$ of the energy transferred to the EM sector~\cite{2020_Cazon_HadInts}. Each neutral pion, indexed by $j$, carries an energy $E_j = E_0 x_j$, in the laboratory frame and immediately decays into two photons. We assume each photon carries half of the energy of the parent $\pi^0$, triggering an EM cascade reaching their maximum at a depth given by Equation~\ref{eq:heitler_xmax}. Therefore, we are neglecting the stochastic fluctuations of the partition of energy between each photon and the fluctuations of the individual photon cascades. Under these assumptions, the depth of the maximum of the cascade triggered by a single neutral pion is
\begin{equation} \label{eq:npion_xmax}
    \lambda_r^{-1}X_{\max, j}^{\pi^0}\equiv t_{\max, j}^{\pi^0} = \ln \left(\frac{E_0}{2 \xi_c^e}\right) + \ln x_j,
\end{equation}
\par\noindent
where we defined the dimensionless depth $t = \lambda_r^{-1}X$, as done in~\cite{1952_Rossi_CEbook}. The first term in Equation~\eqref{eq:npion_xmax} is simply the mean depth of the maximum of an EM cascade triggered by a photon with energy $E_0 / 2$. The second term is always negative and includes the stochastic fluctuations of the fraction of the primary energy carried by neutral pions. For neutral pions with $x_j < 2 \xi_c^e / E_0$, the quantity \( t_{\max, j}^{\pi^0} \) becomes negative, reaching a minimum value of $\lambda_r \ln (M_{\pi^0} / 2 \xi_c^e) \sim - \SI{10}{\depth}$, where $M_{\pi^0}$ is the neutral pion's mass. While these cases are unphysical, Equation~\eqref{eq:npion_xmax} applies only when the decaying pion has enough energy to trigger an EM cascade. Moreover, the impact of such unphysical terms is negligible when considering contributions from other secondary particles of the primary interaction.
\par
The stochastic fluctuations of $x_j$ translate into fluctuations of $t_{\max, j}^{\pi^0}$. The EM cascades triggered by neutral pions develop independently, and their superposition forms the longitudinal profile of the EM cascade triggered by all neutral pions in the first interaction. Instead of modelling each individual profile using Greisen functions~\cite{1960_Greisen_GreisenFunc} and numerically finding the peak of their sum, we estimate the depth of the maximum of the combined electromagnetic (EM) cascade, $t_{\max}^{\text{EM}}$, by computing a weighted average of the individual $t_{\max, j}^{\pi^0}$. The weights are given by the Heitler-model prediction for the maximum number of particles in each sub-cascade, $N_{\max, j} = E_0 x_j / 2\xi_c^e$. Using shower simulations, as described in Section~\ref{sec:xi_model_performance}, we verified that the weighted average provides an unbiased estimate of the position of the maximum of summed Greisen profiles. When the true values of $N_{\max, j}$, which fluctuate for a fixed photon energy, are used as weights, the estimation of $t_{\max}^{\rm EM}$ has an uncertainty of $\SI{5}{\depth}$. If, instead, the approximation $N_{\max, j} \propto x_j$ is used, the uncertainty increases to $\SI{24}{\depth}$.

Letting $m_{\text{EM}}$ denote the number of neutral pions produced in the cosmic ray-air interaction, we obtain
\begin{equation} \label{eq:xi_model_em_first_int}
\begin{aligned}
    t_{\max}^{\mathrm{EM}} & = \frac{\displaystyle \sum_{j = 1}^{\multem} N_{\max, j} t_{\max, j}^{\pi^0}}{\displaystyle \sum_{j = 1}^{\multem} N_{\max, j}} = \ln \left( \frac{E_0}{2 \xi_c^e}\right) + \frac{\displaystyle \sum_{j = 1}^{m_{\text{EM}}} x_j \ln x_j}{\displaystyle \sum_{j = 1}^{m_{\text{EM}}} x_j} = \\
    & = \ln \left( \frac{E_0}{2 \xi_c^e}\right) - \frac{\zetaem}{\alphaem},
\end{aligned}
\end{equation}
\par\noindent
where we identified the fraction of energy carried by all neutral pions of the first interaction $\alphaem \equiv \sum_{j = 1}^{m_{\text{EM}}} x_j$ and defined the multiparticle production variable $\zetaem \equiv - \sum_{j = 1}^{m_{\text{EM}}} x_j \ln x_j$, with $\zetaem > 0$. We discuss the interpretation of $\zetaem$ in terms of its properties in Section~\ref{subsec:new_variables_in_eas}.
\par
Next, we treat the  electromagnetic showers generated by the hadronic sector of the primary interaction, as schematically represented in Figure~\ref{fig:xi_model_scheme}. Suppose each hadronically interacting particle carries a fraction $x_i$ of the primary energy, in the laboratory frame, adding up to the fraction of energy retained in the hadronic sector: $\alphahad$. Moreover, let $m_{\text{had}}$ denote the total number of such particles. The contribution of hadron $i$ to $X_{\max}$ is assessed by considering only neutral pions arising from its interaction with an air atom, neglecting neutral pion decays in subsequent interactions. Moreover, let $\lambda_i$ be the interaction length of hadron $i$, in units of radiation length $\lambda_r$. Under these assumptions and applying Equation~\eqref{eq:xi_model_em_first_int}, the depth of the shower maximum of the cascade started by hadron $i$ is given by
\begin{equation}
    t_{\max, i}^{\text{had}} = \ln \left( \frac{E_0}{2 \xi_c^e}\right) + \lambda_i + \ln x_i - \frac{\zetaem^{(i)}}{\alphaem^{(i)}}.
\end{equation}
\par
The first term is the $\xmax$ for a photon with energy $E_0 / 2$ and the others are stochastic corrections to account for the fraction of the energy of the secondary hadron inducing the EM cascade.
\par
We estimate the combined $X_{\max}$ of the EM cascades triggered by each hadron by averaging all $t_{\max, i}^{\rm{had}}$ values weighted by the number of particles at the shower maximum, $N_{\max, i}$. For hadronic showers, it holds $N_{\max, i} \approx E_i$\footnote{This dependence was verified by fitting the energy evolution of $N_{\max}$ to $y = p_1 \left(E_0/ \xi_c^e\right)^{\gamma} + p_2$, yielding $\gamma = \num{1.004(1)}$, in agreement with~\cite{2005_Matthews_HMmodel, 2012_Kampert_MassModelDep}. The values of $N_{\max}$ were obtained with \conex{} simulations, as described in Section~\ref{sec:xi_model_performance}, for primary energies $E_0 \in [10^{12}, \, 10^{19}[$ eV.}, so that
\begin{equation} \label{eq:xi_model_had_first_int}
    t_{\max}^{\text{had}} = \ln \left( \frac{E_0}{2 \xi_c^e}\right) + \frac{1}{\alphahad} \left[ \sum_{i = 1}^{m_{\text{had}}} x_i \left( \lambda_i - \frac{\zetaem^{(i)}} {\alphaem^{(i)}} \right) - \zetahad \right],
\end{equation}
\par\noindent
where we defined $\zetahad \equiv - \sum_{i = 1}^{m_{\text{had}}} x_i \ln x_i$, with $\zetahad > 0$ in analogy with $\zetaem$.
\par
Finally, we average the depths of the shower maxima obtained in Equations~\eqref{eq:xi_model_em_first_int} and \eqref{eq:xi_model_had_first_int}, pertaining to the EM and hadronic sectors of the primary interaction, respectively, weighting them by the energy contained in each sector
\begin{equation} \label{eq:xi_model_general_form}
\begin{aligned}
    \lambda_r^{-1}\xi & \equiv \alpha_{\text{had}} t_{\max}^{\text{had}} + (1 - \alphahad) t_{\max}^{\text{EM}} = \\
    & = \ln \left( \frac{E_0}{2 \xi_c^e}\right) + \left[ \sum_{i = 1}^{m_{\text{had}}} x_i \left( \lambda_i - \frac{\zetaem^{(i)}}{\alphaem^{(i)}} \right) \right] - \zetahad - \zeta_{\text{EM}}.
\end{aligned}
\end{equation}
\par
Therefore, $\xi$ is the variable of the first interaction that predicts the $\dxmax$ of a given cascade, in a shower-by-shower fashion. The shower-by-shower estimator of $X_{\max}$ is obtained by adding the independent fluctuations of the depth of the first interaction
\begin{equation} \label{eq:xi_model_xmax}
    X_{\max} = X_1 + \xi.
\end{equation}
\par
As defined in Equation~\eqref{eq:xi_model_general_form}, $\xi$ still includes shower-to-shower fluctuations of the interaction depth of each hadronically interacting particle of the first interaction, as well as fluctuations in the particle production upon interaction. We isolate fluctuations solely due to the first interaction using additional approximations:
\begin{enumerate}
\item Parametrise the average value of $\lambda_i$ as a function of incident's hadron energy $E_i^{\text{had}} = E_0 x_i$ as~\cite{2002_Jaime_HybridSims}:
\begin{equation} \label{eq:int_length_parametrisation}
    \lambda_i(E_i^{\text{had}}) = \lambda_0 - \delta \ln \left( \frac{E_0}{E_{\text{ref}}} \right) - \delta \ln x_i,
\end{equation}
\par\noindent
with $\delta > 0$ and where $E_{\text{ref}}$ is some reference energy such that $\lambda_i(E_{\text{ref}}) = \lambda_0$. All quantities are expressed in terms of the radiation length $\lambda_r$ so that $\lambda_0$ and $\delta$ are dimensionless. Note that the interaction length of each hadron depends on its nature and on the hadronic interaction model. To simplify the expression for $\xi$, we neglect the dependence on the nature of the hadron. In Section~\ref{subsec:xi_model_tuning}, we discuss the estimations and tuning of the parameters $\delta$ and $\lambda_0$.

\item Parametrise the average ratio $\zeta_{\text{EM}, i}/ \alpha_{\text{EM}, i}$ as a function of the energy of the parent hadron, $E_i^{\text{had}}$ in the form \footnote{This functional form is motivated by the fact that for equipartition of energy $\zetaem / \alphaem = \ln \multtotal$ and the average $\multtotal$ increases approximately as a power law in energy~\cite{2002_Jaime_HybridSims}}
\begin{equation} \label{eq:zetaem_ratio_parametrisation}
\frac{\zeta_{\text{EM}, i}}{\alpha_{\text{EM}, i}} \simeq \expval{\frac{\zetaem}{\alphaem}} = a_0 + a_1 \ln \left( \frac{E_0 x_i}{E_{\text{ref}}} \right),
\end{equation}
\par\noindent
with $a_1 > 0$. Note that $a_1$ depends on the hadronic interaction model, while $a_0$ can be computed at an energy $E_{\text{ref}}$ such that model differences in $\expval{\zetaem / \alphaem}(E_{\text{ref}})$ are negligible.
\end{enumerate}
Inserting the aforementioned parametrisations into Equation \eqref{eq:xi_model_general_form}, and defining the parameters $\omega = 1 - \delta - a_1$ and $\mathcal{C}_0 = \lambda_0 - a_0$, leads to Equation~(\ref{eq:xi_model_simple}):
\begin{widetext}
\begin{equation} \label{eq:xi_model_simple}
\lambda_r^{-1}\xi = \ln(\frac{E_0}{2 \xi_c^e}) + \left[ \mathcal{C}_0 + (\omega - 1) \ln \left(\frac{E_0}{E_{\text{ref}}}\right) \right] \alpha_{\text{had}} - \omega \zeta_{\text{had}} - \zeta_{\text{EM}}.
\end{equation}
\end{widetext}
\par
Therefore, $\xi$ is a variable built from the energy spectra of secondaries of the first $p$-air interaction alone and that provides an event-by-event estimation of $\dxmax$.
\par
Neglecting the contribution of the rest of the cascade, the probability density functions (PDFs) of $X_{\max}$, $\xi$ and $X_1$ can be related as
\begin{equation} \label{eq:xi_model_xmax_pdf}
    p(X_{\max}) =\int p_{_{X_1}}(X_{\max}-  \xi) p(\xi) \dd{X_1}.
\end{equation}
\par
For simplicity, and when there is no ambiguity, the PDF of a variable $x$ will be denoted by $p(x) \equiv p_X(x)$. Furthermore, we use a notation where variable names stand for their respective PDFs, that is $x \equiv p(x)$, when considering operations between PDFs of random variables. That is, we re-write Equation~\eqref{eq:xi_model_xmax_pdf} as
\begin{equation}
X_{\max} = X_1  \otimes \xi,
\end{equation}
\par\noindent
where $\otimes$ denotes a convolution.
\par
In the case where we take only Equation~\eqref{eq:xi_model_em_first_int}, and assume equipartition of energy among secondaries, the model reduces to the standard Heitler-Matthews prediction for $\dxmax$ given in Equation~\eqref{eq:hm_xmax}.
\par
Finally, note that for perfectly elastic proton-air interactions, $\zetahad = \zetaem = 0$ and $\alphahad = 1$. In turn, in interactions where most of the primary energy is taken by a single neutral pion, we have $\alphahad = \zetahad = \zetaem = 0$, reducing $\xi$ to the $X_{\max}$ value given by the Heitler model for a photon primary with energy $E_0 / 2$.


\section{Testing the causal connection between \boldmath{$\xi$} and \boldmath{$\Delta X_{\max}$}} \label{sec:xi_model_performance}

The causal connection between $\xi$, introduced in Equation~\ref{eq:xi_model_simple} and the shower observable $\dxmax$ is quantified using $10^5$ proton-induced showers simulated with \conex{}~v7.80~\cite{2004_Pierog_conex, 2007_Bergmann_conex} with primary energy $E_0 = 10^{19}\,$eV and zenith angle $\theta = 60^\circ$, using the high-energy hadronic interaction models \eposr{}~\cite{2023_Tanguy_eposlhcr}, \qIII{}~\cite{2024_Ostapchenko_qgsIII} and \sibe{}~\cite{2020_Felix_sibyll23d}. Particles above $E_{\text{th}} = 0.005 \times E_0$ and their interactions were tracked individually. Below this value, the longitudinal shower profile was computed by numerically solving cascade equations. The ground level was set to $1\,400\,$m a.s.l, the average height of the Pierre~Auger~Observatory~\cite{2015_Auger_PAODescription}, corresponding to an average vertical depth of $X_{\text{gr}} = 880\,\mathrm{g\,cm^{-2}}$. The value of $\xmax$ is taken from a Gaisser-Hillas fit to the longitudinal profile of all charged particles.

\subsection{Optimization of the free parameters of $\xi$} \label{subsec:xi_model_tuning}

The parameterizations expressed in Equations~\eqref{eq:int_length_parametrisation} and \eqref{eq:zetaem_ratio_parametrisation} led to the introduction of the parameters $\mathcal{C}_0$ and $\omega$ in Equation~\eqref{eq:xi_model_simple}. These parameters do not vary from shower-to-shower, and their values change the relative contribution of the variables $\zetahad$, $\zetaem$ and $\alphahad$ to $\xi$.
\par
By constraining the energy dependencies of the total particle-air interaction cross-sections and of $\expval{\zetaem / \alphaem}$, the parameters $\mathcal{C}_0$ and $\omega$ can be tuned, similarly to what is done in~\cite{2011_Ulrich_f19}. This could be achieved with independent measurements or by measuring the primary interaction as a function of the cosmic ray energy, using, for example, the multiparticle production variables we derive here, thus allowing for a recursive tuning of these free parameters.
\par
Instead, in this work, the values of $\mathcal{C}_0$ and $\omega$ were calculated in two steps:
\begin{enumerate}
    \item by minimising, for each hadronic interaction model, a $\chi^2$ defined by
\begin{equation} \label{eq:free_params_chi2}
    \chi^2 \equiv \sum_{k = 1}^{N} \frac{\left(\Delta X_{\max, k} -   \xi_{ k}\right)^2}{\Delta X_{\max, k}},
\end{equation}
\par\noindent
where $k$ runs over the number of simulated showers, $N$. The obtained value of $\mathcal{C}_0$ is then averaged over the hadronic interaction models, under the assumption that the models converge at a certain energy $E_{\text{ref}}$.
\item Fixing $\mathcal{C}_0$ to its model-averaged value, the $\chi^2$ minimization is repeated with
\begin{equation}
\omega = p_1 \expval{\frac{\zetaem}{1- \alphahad}} + p_0,
\end{equation}
\par\noindent
where the parameters $p_0$ and $p_1$ are simultaneously optimized for the three hadronic interaction models. Therefore, the dependence on the hadronic interaction model is captured by $\expval{\zetaem / (1- \alphahad)}$.
\end{enumerate}
Using the optimal values of $\mathcal{C}_0$, $p_0$ and $p_1$, Equation~\eqref{eq:xi_model_simple} can be written for $E_0 = 10^{19}$ eV as
\begin{widetext}
\begin{equation} \label{eq:xi_model_numerical}
\xi \simeq 917 + \left[ 73 - 48 \expval{\frac{\zetaem}{1 - \alphahad}}\right] \alpha_{\text{had}} + \left[ 5 \expval{\frac{\zetaem}{1 - \alphahad}} - 40 \right] \zeta_{\text{had}} - 37 \times \zeta_{\text{EM}}\quad[\unit{\depth}].
\end{equation}
\end{widetext}

\subsection{Evaluating the causal connection between \boldmath{$\xi$} and \boldmath{$\dxmax$}} \label{subsec:xi_model_performance}
\begin{figure}[h!]
    \centering
    \includegraphics[width=\columnwidth]{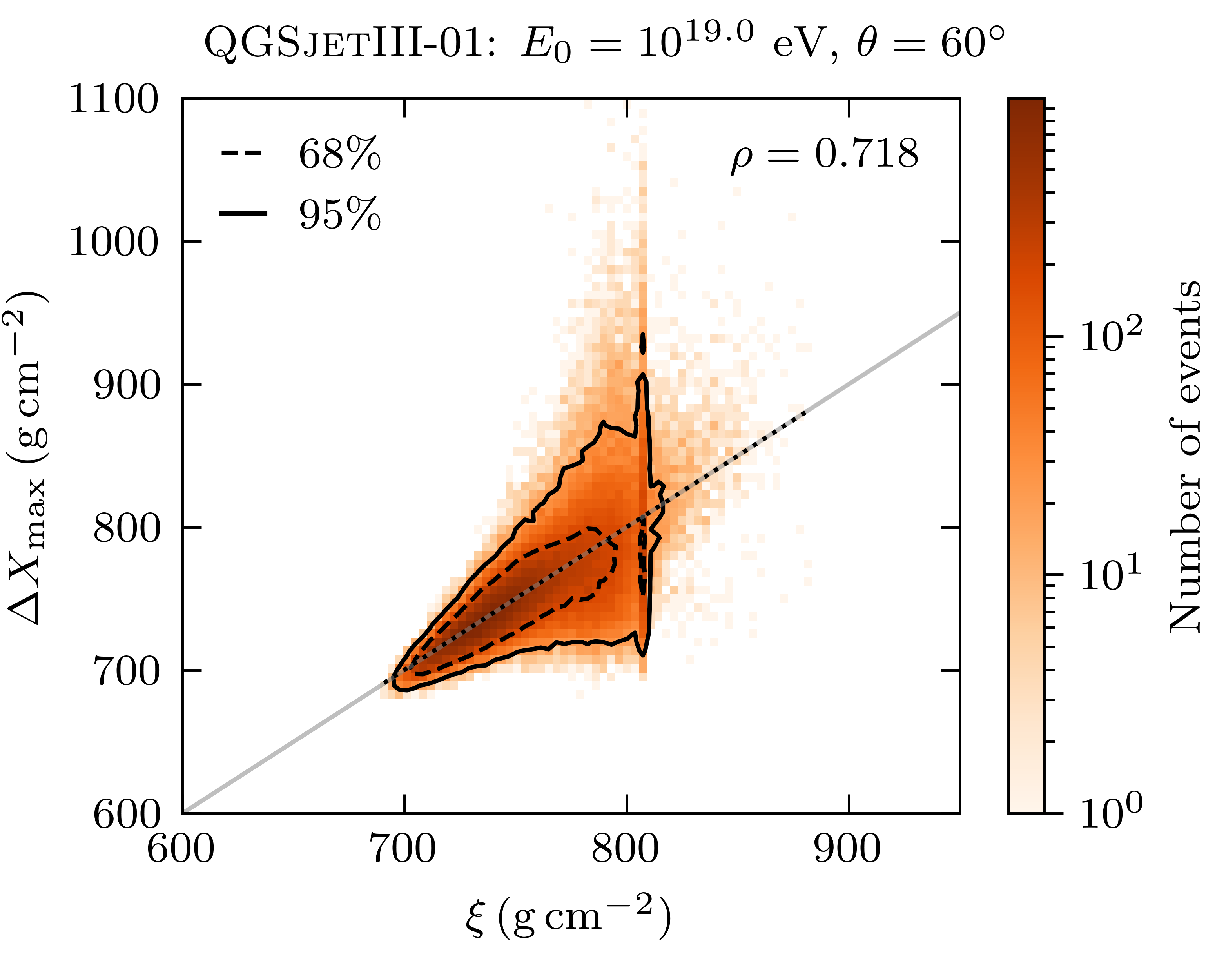}
    \caption{Correlation between the predictor of $\dxmax$ from the first $p$-air interaction, $\xi$, and the event-by-event values of $\dxmax$. The contours containing 68 \% and 95 \% of the events are represented by the dashed and solid black lines, respectively. The 1:1 line is represented in solid grey and the linear regression curve as a dotted black line. This figure was produced using the library of proton-induced \conex{} simulations described in Section~\ref{sec:xi_model_performance}, with the high-energy hadronic interaction model \qIII{}.}
    \label{fig:simpleXi_vs_xmax_eposlhc}
\end{figure}

The strong connection between the shower-to-shower values of $\xi$ and $\dxmax$ is evident from Figure~\ref{fig:simpleXi_vs_xmax_eposlhc}, which displays the correlation of these quantities for the high-energy hadronic interaction model \qIII{}, together with their Pearson correlation coefficient. The dashed and solid black contours represent the boundaries containing $68 \%$ and $95 \%$ of the events in the sample. In addition, the figure displays the 1:1 line (solid grey) and a curve (dotted black) resulting from a linear regression to the pairs of points $(  \xi, \Delta X_{\max})$.
\par
\par
The stochastic fluctuations of the primary interaction variable $\xi$ are very well correlated with those of the shower observable $\dxmax$, as demonstrated by their Pearson correlation coefficient $\rho = 0.72$. For reference, using more usual production variables such as the total multiplicity, $\multtotal$, and the elasticity, $\elasticity$, combined as $\dxmax \simeq \ln (E_0 /2 \xi_c^e) + \lambda_r \ln (\elasticity / \multtotal)$~\cite{2012_Kampert_MassModelDep} yields only $0.55 < \rho < 0.63$. This justifies our more careful treatment of the energy spectra of secondaries of the primary interaction. Furthermore, the variance in $\xi$ accounts for $52 \%$ of the variance in $\dxmax$. Therefore, fluctuations in particle production and interaction depth in later shower generations play a sub-dominant role in determining the shower-to-shower value of $\dxmax$. Moreover, the regression line coincides with the 1:1 line in the $(  \xi, \Delta X_{\max})$ plane. Two additional features are clear from Figure~\ref{fig:simpleXi_vs_xmax_eposlhc}. The first is the vertical line passing through $\sim 815\,\mathrm{g\,cm^{-2}}$ which corresponds to diffractive events. The value of $\xi$ corresponding to this line is obtained by inserting $\alphahad = 1$ and $\zetahad = \zetaem = 0$ into Equation~\eqref{eq:xi_model_simple}. The second is the increasing dispersion in $\dxmax$ with $\xi$. This is due to the increase in the elasticity of the primary interaction as explored in Appendix~\ref{apx:leading_fluctuations}.
\par
We have verified that the correlation between $\xi$ and $\Delta X_{\max}$ has no dependence on the zenith angle of the shower, as expected due to the fast decoupling of the electromagnetic component of the shower from its hadronic development. However, their correlation coefficient decreases monotonically with the primary energy down to $\rho \sim 0.65$ at $E_0 = 10^{17}\,$eV. This is due to a lower multiplicity of particles produced in the primary interaction at lower energies, suppressing less the fluctuations in the subsequent shower generations. The greater importance of the stochasticity in later shower generations degrades the correlation between $\xi$ and $\dxmax$.
\par
\begin{figure}[h!]
    \centering
    \includegraphics[width=\columnwidth]{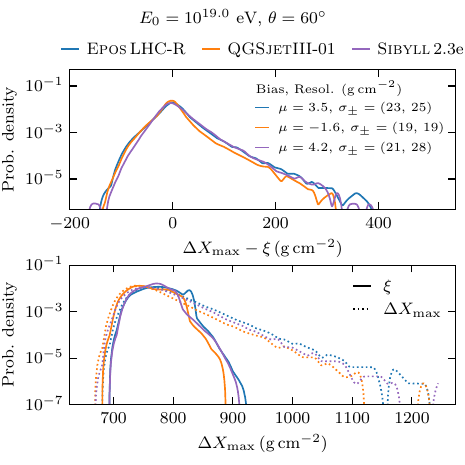}
    \caption{Upper panel: distribution of the residuals $\Delta X_{\max} -   \xi$, along with the bias and resolution in the determination of $\Delta X_{\max}$. Lower panel: distributions of $\Delta X_{\max}$ (dotted lines) and its predictor from the first interaction $  \xi$ (solid lines). The hadronic interaction models \eposr{}, \qIII{} and \sibe{} are represented in blue, orange and purple, respectively. This figure was produced with the library of proton-induced \conex{} described in Section~\ref{sec:xi_model_performance}.}
    \label{fig:simpleXi_residuals_all_models}
\end{figure}

The fluctuations in the development of the cascade after the first interaction and the approximations used in the derivation of the functional form of $\xi$ are included in the residuals $R_X \equiv \Delta X_{\max} - \xi$, whose marginalized distribution, $p(R_X) = \int p(R_X, \xi) \dd{\xi}$ is shown in the upper panel of Figure~\ref{fig:simpleXi_residuals_all_models}, for the high-energy hadronic interaction models \eposr{}, \qIII{} and \sibe{}. The distributions of $\dxmax$ and $\xi$ are shown in the lower panel of the same figure.
\par
The fluctuations of $\Delta X_{\max}$, $\xi$, the Pearson correlation coefficient between $\Delta X_{\max}$ and $\xi$, and the bias and resolution in the prediction of $\Delta X_{\max}$ from $\xi$, are shown in Table~\ref{tab:performance_of_xi_model} for the three hadronic interaction models \eposr{}, \qIII{} and \sibe{}. The strong causal connection between $\xi$ and $\Delta X_{\max}$ depends little on the hadronic interaction model.
\par
\begin{table}[h!]
    \caption{Standard deviation of $\Delta X_{\max}$ and $\xi$, Pearson correlation coefficient between $\Delta X_{\max}$ and $\xi$, bias ($\expval{\Delta X_{\max} -  \xi})$ and left and right resolutions $\sigma_\pm(\dxmax - \xi$), for three hadronic interaction models.}
    \centering
    \begin{tabular}{c | c | c | c}
    \hline \hline
    Parameter & \eposr{} & \qIII{} & \sibe{} \\ \hline
    $\sigma(\Delta X_{\max})$ & 45.1 & 40.1 & 43.0 \\
    $ \sigma(  \xi)$ & 30.3 & 29.3 & 26.6\\
    $\rho$ & 0.68 & 0.72 & 0.68 \\
    $\expval{\Delta X_{\max} -  \xi}$ & 3.5	&  -1.6 & 4.7 \\
    $\sigma_\pm(\Delta X_{\max}-  \xi)$ & (22.7, 25.3) & (19.1, 18.9) & (21.1, 28.5)\\ \hline \hline
    \end{tabular}
    \label{tab:performance_of_xi_model}
\end{table}
\par
The value of $\xi$ for the primary $p$-air interaction estimates the $\dxmax$ of the corresponding cascade with a bias between $\sim \SI{-2}{\depth}$ and $\sim \SI{5}{\depth}$, depending on the hadronic interaction model. The precision in the determination of $\Delta X_{\max}$ varies from $\sigma_\pm \simeq \SI{19}{\depth}$ to $\sigma_\pm \simeq \SI{25}{\depth}$, also depending on the hadronic interaction model. The shape of $p(R_X)$ is identical for all models, displaying an exponential tail towards positive values. The events populating this tail also populate the exponential deep tail of the distribution of $\dxmax$, displayed in the lower panel of Figure~\ref{fig:simpleXi_residuals_all_models}. They correspond to highly elastic primary interactions in which the fluctuations of the interaction depth of the leading particle affect greatly the value of $\dxmax$. This is discussed in detail in Appendix~\ref{apx:leading_fluctuations}, where a modification of $\xi$ is introduced to account for these extra fluctuations and reproduce the tail in the true distribution of $\Delta X_{\max}$. Nevertheless, the agreement of the distributions of $\dxmax$ and $\xi$ for rapidly developing showers is remarkable.
\par
\begin{figure}[h!]
    \centering
    \includegraphics[width=\columnwidth]{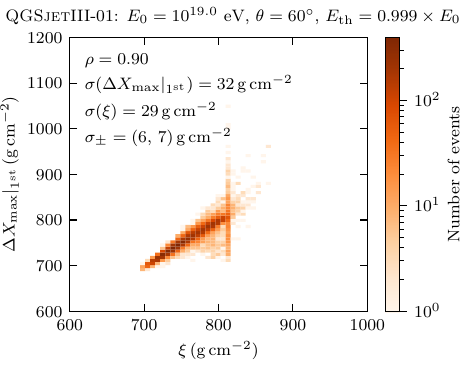}
    \caption{Correlation plot between $  \xi$ and $\Delta X_{\max}|_{1^{\text{st}}}$, for stochastic first interactions and a deterministic prediction of the rest of the shower. The distribution is obtained with \conex{} proton-induced simulations as described in the main text, but with the energy threshold between Monte-Carlo and Cascade Equations set to $E_{\text{th}} = 0.999 \times E_0$, using the high-energy hadronic interaction model \qIII{}.}
    \label{fig:simpleXi_vs_dXmax_frozen_shower}
\end{figure}

The fluctuations in particle production and propagation after the primary interaction and the approximations used in the derivation of $\xi$ contribute to the variance of the residuals $R_X$. These separate contributions are assessed using $10^4$ proton-induced \conex{} simulations with the threshold from Monte-Carlo to cascade equations set to $E_{\text{th}} = 0.999 \times E_0$. In this setting, the stochastic fluctuations in the shower development are confined to the primary interaction, as the rest of the shower develops deterministically. In this setting, the resulting event-by-event values of $\Delta X_{\max} |_{1^{\text{st}}}$ against $\xi$ is shown in Figure~\ref{fig:simpleXi_vs_dXmax_frozen_shower}.
\par
\par
The Pearson correlation coefficient of $\rho = 0.90$ shows that $\xi$ captures the majority of the variability in $\dxmax$ due to the stochasticity of the primary interaction alone.
The amount of information lost by the approximations taken in the derivation of $\xi$ is measured by the standard deviation of $\Delta X_{\max}|_{1^{\text{st}}} - \xi$, which amounts to $\sigma(\Delta X_{\max}|_{1^{\text{st}}} - \xi)= \SI{14}{\depth}$. 
The fluctuations of $\dxmax$ for a fully stochastic cascade can be decomposed as
\begin{equation}
    \sigma^2(\Delta X_{\max}) \simeq \sigma^2(\xi) + \sigma^2(\Delta X_{\max}|_{1^{\text{st}}}- \xi) + \sigma_R^2,
\end{equation}
\par\noindent
where $\sigma^2_R$ denotes the variance in $\dxmax$ due to fluctuations in all shower generations but the first, $\sigma^2(\xi)$ is the variance of $\xi$ and $\sigma^2(\Delta X_{\max}|_{1^{\text{st}}}- \xi)$ the variance due to the approximations in the derivation of $\xi$. The relative contributions of each term are $\sigma^2(\xi) = 54 \%$, $\sigma^2(\Delta X_{\max}|_{1^{\text{st}}}- \xi) = 12.6 \%$ and $\sigma^2_R = 33.4 \%$. Therefore, $66.6 \%$ of the variance in $\dxmax$ is attributed to the stochasticity of the primary interaction. Furthermore, $\xi$ captures $54 / 66.6 \simeq 80\%$ of the variability of the primary interaction.
\par
In summation, $\sim 80 \%$ of the fluctuations in the primary interaction that determine the variance in $\dxmax$ are explained by the variance in $\xi$. Fluctuations in later shower generations are such that, for a fully stochastic cascade, the variance in $\xi$ explains $54\%$ of the variance in $\dxmax$. The degradation of the causal connection between $\xi$ and $\dxmax$ is mostly due to variability in interactions in later shower generations and not due to a breakdown of the assumptions used in the derivation of $\xi$. One exception is the case where an extremely energetic neutral pion is produced in the first $p$-air interaction. In this case, $\alphahad = \zetahad = 0$ and $\zetaem \to 0$ yielding a fixed value of $\xi$ despite the fluctuations of the maxima of the two resulting electromagnetic showers.
\subsection{Energy evolution of the correspondence between \boldmath{$\xi$} and \boldmath{$\dxmax$}}
We further validate the connection between $\xi$ and $\dxmax$ by computing the primary energy dependence of the first moment of $\xi$ in the energy range $\log_{10} (E_0 / \mathrm{ eV}) = [17, \,20]$. For all primary energies, the parameter $\mathcal{C}_0$ in Equation~\eqref{eq:xi_model_simple} is fixed to its value fitted at $E_0 = 10^{19}\,$eV, while $\omega$ is generalized as
\begin{equation} \label{eq:omega_energy_func}
    \omega(E_0) = p_1(E_0) \expval{\zetaem / (1 - \alphahad)} + p_2(E_0),
\end{equation}
\par\noindent
where $p_1(E_0)$ and $p_2(E_0)$ depend on $E_0$ as indicated in Table~\ref{tab:xi_params_energy_func} in Appendix~\ref{apx:xi_params_energy_func}. The dependence on the hadronic interaction model is encapsulated in $\expval{\zetaem / (1 - \alphahad)}$.
\par
\begin{figure}[h!]
    \centering
    \includegraphics[width=\columnwidth]{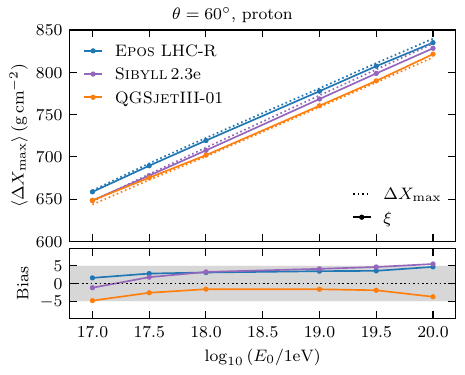}
    \caption{Top panel: Average values of $\xi$ and $\dxmax$ as a function of the primary energy. Bottom panel: Bias $\expval{\dxmax - \xi}$ as a function of the primary energy, for the hadronic interaction models \eposr{}, \qIII{} and \sibe{}.}
    \label{fig:xi_dXmax_elongation_rate_all_models}
\end{figure}

The elongation rates of $\xi$ and $\dxmax$ are shown in Figure~\ref{fig:xi_dXmax_elongation_rate_all_models}, for the high energy hadronic interaction models \eposr{}, \qIII{} and \sibe{}, as dotted lines.
\par
\par
The evolution of $\expval{\xi}$ is linear with the logarithm of the primary energy, reproducing the expected energy dependence of $\expval{\dxmax}$, with biases smaller than $\pm\SI{5}{\depth}$ across three decades in primary energy. This validates the parameterizations employed in the definition of $\xi$ as well as the approximations taken in the derivation of this quantity. Additionally, the predicted elongation rate from $\xi$ has a bias of, at most, $\SI{2}{\depth}$ depending on the high-energy hadronic interaction model.


\section{A complete and universal model of the distribution of \boldmath{$X_{\max}$}} \label{sec:univerality}
\subsection{Universality of the shower response to $\xi$}
We have shown that the fluctuations of $\xi$ determine most of the fluctuations of $\dxmax$. The remaining variability, which arises from hadron production and propagation in later stages of shower development, is captured by the residuals $R_X = \dxmax - \xi$. The probability density function of these residuals, denoted by $p(R_X\,|\,\xi, M)$, encodes the response of the rest of the shower to a given primary-interaction value of $\xi$ for a specific hadronic interaction model $M$. For brevity, we henceforth omit the explicit dependence of the PDF of the residuals on $\xi$ and simply write $p(R_X\,|\,M)$. The probability density function $p(\dxmax)$ can be cast as the integral transformation or convolution
\begin{equation}
p(\dxmax) = \int p(\xi) p(R_X\,|\, M)\dd{\xi},
\end{equation}
\par\noindent
where $p(\xi)$ is the PDF of $\xi$, and $p(R_X\,|\,M)$ is the kernel of the transformation. This result follows trivially by noting that $\dxmax = \xi + R_X$, and can be written in the simplified notation
\begin{equation} \label{eq:shower_response}
\dxmax = \xi \otimes R_X\,|\,M,
\end{equation}
where the symbol $\otimes R_X\,|\,M$ denotes convolution with the kernel $p(\dxmax - \xi\,|\,M)$. The results of Section~\ref{subsec:xi_model_performance}, show that the kernel is narrow compared to $p(\xi)$.
\par
Additionally, the possibility of reconstructing $p(\xi)$ from a measured distribution of $\xmax$ also depends on how the shower response to a fixed value of $\xi$ varies with the hadronic interaction model $M$. That is, on the universality of the kernel, $p(R_X\,|\,M)$. Using the library of \conex{} shower simulations described in Section~\ref{sec:xi_model_performance}, we assess this degree of universality. To do so, we first define an average kernel, $\overline{p(R_X)}$, obtained by re-centering the individual kernels $p(R_X \mid M)$ of each hadronic interaction model, and then averaging them over $M \in$ \{\eposr{}, \qIII{}, \sibe{}\}. For each hadronic interaction model, the average shower response can then be used to map $p(\xi)$ into $p(\Delta X_{\max})$ as $\overline{\dxmax} = \xi \otimes \overline{R_X}$.
\par
\begin{figure}[ht!]
    \centering
    \includegraphics[width=\linewidth]{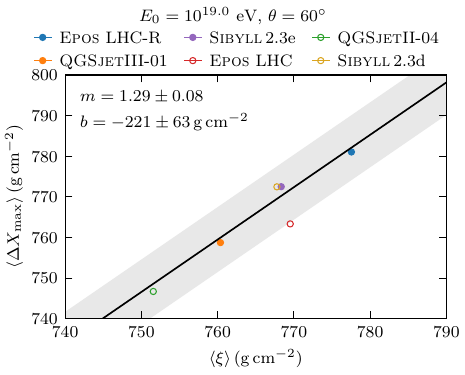}
    \caption{Values of $\expval{\xi}$ against $\expval{\dxmax}$ for the hadronic interaction models \eposr{}, \qIII{} and \sibe{}, \epos{}, \qII{} and \sibd{}. The most up-to-date hadronic interaction models are depicted with full circles, while their previous versions are represented by the empty markers. The black straight line results from a linear regression to the pairs of points $(\xi, \dxmax)$ produced with the most-up-to-date models. The $\pm$\SI{8}{\depth} band is represented in light-grey This figure was produced with the library of proton-induced \conex{} described in Section~\ref{sec:xi_model_performance}.}
    \label{fig:mean_xi_calibration}
\end{figure}

However, allowing for changes in particle production in the primary interaction of the shower, while keeping all subsequent interactions fixed to an average shower response $\overline{p(R_X)}$, would require abrupt variations in multiparticle production properties as a function of interaction energy. In the absence of new physics, this is highly challenging. Thus, and since energy partition fluctuations diminish in later shower generations, we test whether hadronic interaction models propagate changes from $\expval{\xi}$ to $\expval{\dxmax}$ in a similar manner. This is shown in Figure~\ref{fig:mean_xi_calibration}. By applying a linear regression, $\expval{\dxmax} = m \expval{\xi} + b$, we can shift the value of $\overline{\dxmax}$ for each prior $p(\xi)$, to ensure that $\expval{\overline{\dxmax}} = m \expval{\xi} + b$, effectively propagating spectral changes through the shower. Therefore, the shower response is made explicitly dependent on $\expval{\xi}$, rather than on the hadronic interaction model $M$.
\par
\par
This approach is validated by including other high-energy hadronic interaction models \epos{}~\cite{2015_Pierog_eposlhc}, \qII{}~\cite{2011_Ostapchenko_qgsjet} and \sibd{}~\cite{2020_Felix_sibyll23d} in Figure~\ref{fig:mean_xi_calibration}. For these models, the maximum deviance with respect to the $\xi$-$\dxmax$ calibration curve is $\SI{7}{\depth}$.
\par
We thus define a new response function $p(R_X | \expval{\xi})$, which is solely dependent on $p(\xi)$ and $\expval{\xi}$, and therefore independent of the hadronic interaction model. The final model for predicting $p(\dxmax)$ for any prior $p(\xi)$ is thus given by
\begin{equation} \label{eq:dxmax_complete_model}
    \dxmax = \xi \otimes R_X | \expval{\xi}.
\end{equation}
\par
\begin{figure*}[ht!]
    \centering
    \includegraphics[width= \linewidth]{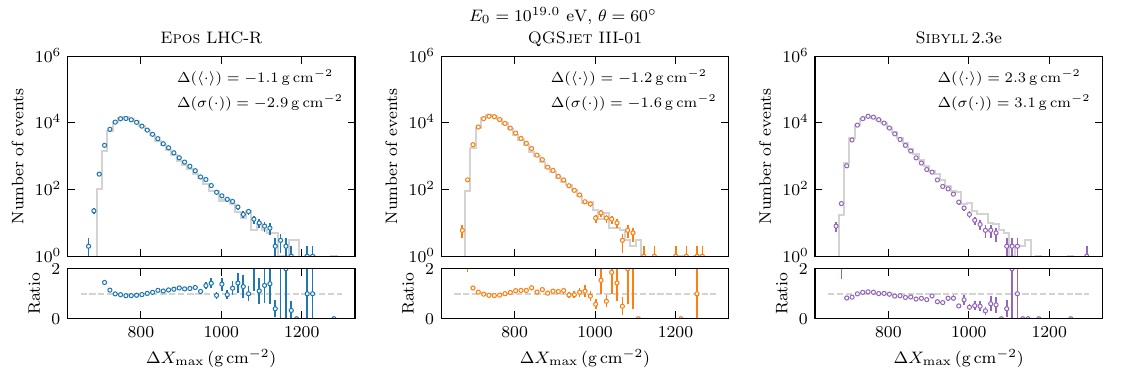}
    \caption{True (grey steps) and predicted (coloured markers) distributions of $\dxmax$ using an average kernel, for distributions of $\xi$ produced with the hadronic interaction models \eposr{} (left), \qIII{} (middle) and \sibe{} (right). This figure was produced with the library of proton-induced \conex{} described in Section~\ref{sec:xi_model_performance}.}
    \label{fig:dxmax_rec_average_response}
\end{figure*}

We validate this model by generating prior distributions of $\xi$ using the hadronic interaction models \eposr{}, \qIII{} and \sibe{}, and predicting the corresponding distributions of $\dxmax$ using Equation~\eqref{eq:dxmax_complete_model}. These are shown in Figure~\ref{fig:dxmax_rec_average_response} together with the true distributions of $\dxmax$ for each hadronic interaction model. The figure shows biases in the predicted mean and standard deviation of $\dxmax$. These biases are denoted by $\Delta(\expval{\dxmax})$ and $\Delta(\sigma(\dxmax))$, and their spread across hadronic interaction models is taken as a systematic uncertainty. The bottom panel displays the ratio of the predicted to the true $\dxmax$ distribution.
\par
\par
The true $\Delta X_{\max}$ distribution is well reproduced except for the most extreme values. The model systematic uncertainty in the mean $\Delta X_{\max}$ is below $\sim \SI{2.5}{\depth}$, which corresponds to about $13\%$ of the inter-model difference in $\expval{\Delta X_{\max}}$. This level of precision allows for model discrimination. However, the systematic uncertainty in $\sigma(\dxmax)$ reaches $\SI{3}{\depth}$, which comparable to the actual differences between hadronic interaction models.
\subsection{Prediction of $X_{\max}$ from $\xi$}
With the approach defined in the previous section, we estimate the distribution of $X_{\max}$, $p(X_{\max})$, as
\begin{equation} \label{eq:xmax_complete_model}
X_{\max} = \xi \otimes R_X|\expval{\xi} \otimes X_1.
\end{equation}
\par
\begin{figure*}[th!]
    \centering
    \includegraphics[width=\linewidth]{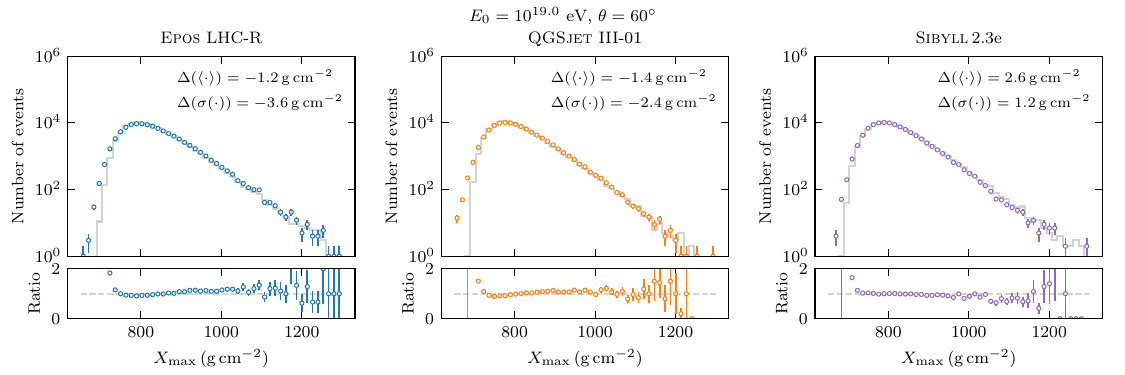}
    \caption{True (grey steps) and predicted (coloured markers) distributions of $\xmax$ using an average kernel, for distributions of $\xi$ produced with the hadronic interaction models \eposr{} (left), \qIII{} (middle) and \sibe{} (right). This figure was produced with the library of proton-induced \conex{} described in Section~\ref{sec:xi_model_performance}.}
    \label{fig:simpleXI_xmax_distributions_all_models}
\end{figure*}

The latter formulation also averages the differences in proton-air cross-section between the models and includes the calibration between $\expval{\xmax}$ and $\expval{\xi}$, as discussed in the previous section. This introduces a systematic of $\sim \SI{3}{\depth}$. The true and estimated distributions of $\xmax$ using Equation~\eqref{eq:xmax_complete_model} are shown in Figure~\ref{fig:simpleXI_xmax_distributions_all_models}, for the hadronic interaction models \eposr{}, \qIII{} and \sibe{}.
\par
\par
The overall agreement between the true and predicted $X_{\max}$ distributions is remarkable and demonstrates the universality of the correspondence between $\xi$ and $X_{\max}$. Table~\ref{tab:xi_xmax_moments} shows the first and second moments of the distributions of $X_{\max}$ (true) and $\overline{X_{\max}}$ (predicted), together with the values of the slope of the right tails of both distributions, $\Lambda_\eta$. As investigated in~\cite{2012_Auger_xsection}, $\Lambda_\eta$ is extracted from an unbinned likelihood fit to the $20 \%$ deepest events in the tail of $X_{\max}$ using an exponential function of the form $y \propto \exp{- X_{\max} / \Lambda_\eta}$.
\par
\begin{table}[h!]
    \caption{Moments of the true distribution of $X_{\max}$ and for its estimation $\overline{\xmax} = X_1  \otimes \xi \otimes R_X|\expval{\xi}$, for the different hadronic interaction models. The definition of $\Lambda_\eta$ is given in the main text. This table was produced with the library of proton-induced \conex{} described in Section~\ref{sec:xi_model_performance}.}
    \centering
    \begin{tabular}{c | c | c | c}
    \hline \hline
    Moment & \eposr{} & \qIII{} & \sibe{} \\ \hline
    $\expval{X_{\max}}$ & 824.44 & 803.37 & 817.11 \\
    $\sigma(X_{\max})$ & 62.10 & 59.29 & 61.58 \\
    $\Lambda_\eta(X_{\max})$ & \num{52.48(37)} & \num{51.56(36)} & \num{51.88(37)} \\ \hline
    $\expval{\overline{X_{\max}}}$ & 825.67 & 804.78 & 814.47 \\
    $\sigma(\overline{X_{\max}})$ & 65.68 & 61.69 & 60.42 \\
    $\Lambda_\eta(\overline{X_{\max}})$ & \num{53.68(38)} & \num{51.46(36)} & \num{50.17(35)} \\ \hline \hline
    \end{tabular}
    \label{tab:xi_xmax_moments}
\end{table}
\par
The systematic uncertainty in estimating $\expval{X_{\max}}$ is $\SI{2.5}{\depth}$, about 10 \% of the difference between hadronic interaction models, allowing for their discrimination solely based on differences in energy spectrum of secondaries of the primary interaction. The biases in estimating $\sigma(X_{\max})$ and $\Lambda_\eta$ are \SI{3}{\depth} and \SI{2}{\depth}, respectively. The strong causal link between $\xi$ and $X_{\max}$, along with the universality of the shower response to $p(\xi)$, ensures that a measured $\xmax$ distribution can constrain the energy spectra of secondary hadrons in kinematic regimes beyond the reach of human-made colliders.
\section{New multiparticle production variables in extensive air showers} \label{sec:new_variables}
\subsection{Interpretation of the new multiparticle production variables} \label{subsec:new_variables_in_eas}
The derivation of the functional form of $\xi$ led to the introduction of new multiparticle production variables, namely $\zetahad$ and $\zetaem$, defined by
\begin{equation}
    \zetahad \equiv - \sum_{i = 1}^{\multhad} x_i \ln x_i \quad ; \quad \zetaem \equiv - \sum_{j = 1}^{\multem} x_j \ln x_j.
\end{equation}
\par
These variables quantify the contribution of each secondary of the primary interaction to the overall electromagnetic longitudinal profile of the induced cascade. More insight is gained by looking at kinematic bounds on $\zetaem$ and $\zetahad$. For each fixed value of $\alphahad$, we maximise the values of these quantities, under the restriction of energy conservation, to yield the bounds
\begin{equation} \label{eq:zeta_kin_bounds}
\begin{aligned}
    - \alphahad \ln \alphahad & \leq \zetahad \leq \alphahad \ln \left( \frac{\multhad}{\alphahad}\right) \\
    \quad - \alphaem \ln \alphaem & \leq \zetaem \leq \alphaem \ln \left( \frac{\multem}{\alphaem}\right)
\end{aligned}
.
\end{equation}
\par
The lower bounds correspond to $\multhad = 1$ and $\multem = 1$, in the respective sector. The upper bounds correspond to the equipartition of the primary energy within each particle sector for each fixed number of secondary particles. If energy is equipartitioned between all secondary particles, irrespectively of the particle sector, we get
\begin{equation}
    0 \leq \zetaem + \zetahad \leq \ln \multtotal.
\end{equation}
\par
Between these bounds, $\zetahad$ and $\zetaem$ are sensitive to the multiplicity and asymmetry in the energy sharing between secondary particles. In particular, when the incident proton scatters elastically off the air target: $\alphahad = \multhad = 1$ and $\multem = 0$, resulting in $\zetaem = \zetahad = 0$.
\par
\begin{figure*}[ht!]
    \centering
    \includegraphics[width= \linewidth]{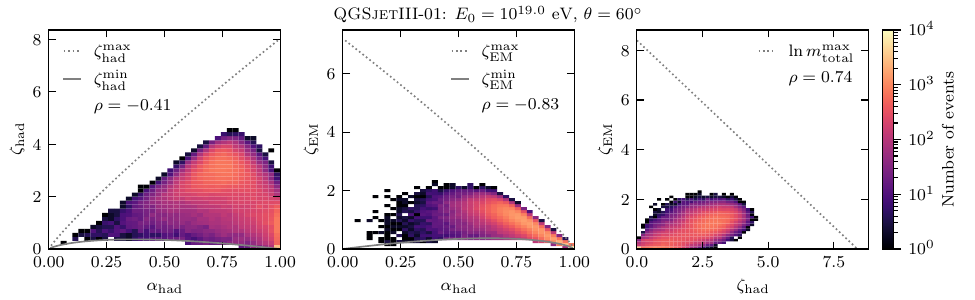}
    \caption{Joint distributions of $\zetahad$ and $\alphahad$ (left panel), $\zetaem$ and $\alphahad$ (middle panel) and $\zetahad$ and $\zetaem$ (right panel), together with kinematic limits imposed by conservation of energy. This figure was produced with the library of proton-induced \conex{} described in Section~\ref{sec:xi_model_performance}, using \qIII{}.}
    \label{fig:zeta_alpha_joint_dists}
\end{figure*}

The joint distributions $f(\zetahad, \alphahad)$, $f(\zetaem, \alphahad)$ and $f(\zetahad, \zetaem)$ are shown in Figure~\ref{fig:zeta_alpha_joint_dists} for the high-energy hadronic interaction model \qIII{}, together with dashed grey lines representing the bounds expressed in Equation~\eqref{eq:zeta_kin_bounds}. The upper bounds on $\zetahad$ and $\zetaem$ depend on the multiplicity of each event, so the limit was derived by taking the maximum multiplicity of hadronic and electromagnetic particles over the ensemble of showers. We thus defined $\zetahad^{\min} \equiv - \alphahad \ln \alphahad$ and $\zetahad^{\max} \equiv \alphahad \ln \left( \multhad^{\max} / \alphahad \right)$, and analogously for $\zetaem$.
\par
\par
The bounds shaping the aforementioned joint distributions are further determined by the elasticity of the primary interaction and by the correlation between multiplicity and the total fraction of energy carried by the particles in each sector.
\par
\begin{figure*}[ht!]
    \centering
    \includegraphics[width= \linewidth]{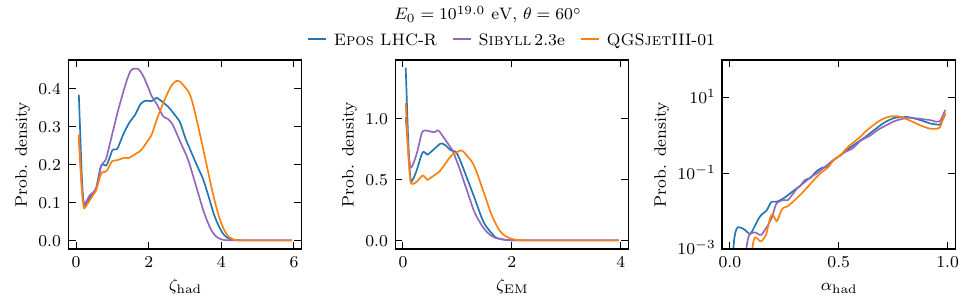}
    \caption{Distributions of $\zetahad$ (left panel), $\zetaem$ (middle) panel and $\alphahad$ (right panel), for different hadronic interaction models \eposr{}, \qIII{} and \sibe{}. This figure was produced with the library of proton-induced \conex{} described in Section~\ref{sec:xi_model_performance}.}
    \label{fig:zeta_dists}
\end{figure*}

The distributions of $\zetahad$, $\zetaem$ and $\alphahad$ over the ensemble of \conex{} simulations described in Section~\ref{sec:xi_model_performance} are shown in Figure~\ref{fig:zeta_dists}, for the hadronic interaction models \eposr{}, \qIII{} and \sibe{}.
\par
\par
The distribution of $\zetahad$ is the $\multhad$-fold weighted convolution of the energy spectra of hadronically interacting particles of the primary interaction. Larger values of $\zetahad$ characterize primary interactions with higher hadronic activity. In these interactions, there is a large number of produced secondaries, so that the probability density function of $\zetahad$ becomes Gaussian in the large $\zetahad$ limit. Applying the same reasoning to the electromagnetic sector yields the same conclusions for $\zetaem$. In turn, the distributions of $\zetahad$ and $\zetaem$ towards lower values are populated with more elastic first interactions. In particular, the peaks at $\zetahad = \zetaem = 0$ are due \textit{quasi}-elastic proton-air scattering.
\par
Crucially, the shapes of the distributions of $\zetahad$ and $\zetaem$ are highly dependent on the hadronic interaction model. This shows that these variables are highly sensitive to the particular physical mechanisms and parametrisations employed by each hadronic interaction model. Therefore, constraining the shape of these distributions allows for strong model discrimination.
The exact relation between the shapes of the distributions of these variables and the shape of the energy spectra of secondary particles is under study. The features of distribution of $\alphahad$ were thoroughly discussed in~\cite{2018_Cazon_alpha, 2021_Cazon_lambdamu, 2024_Martins_lambdamu_xmax}.
\subsection{Interpretation of new production variables in terms of kinematic variables}
We can also cast the variables $\zetaem$ and $\zetahad$ in terms of the pseudo-rapidity $\eta_i$ of each secondary particle $i$. If $\vb{p}_i$ is the momentum of particle $i$ and $p_{\parallel, i}$ its longitudinal component, then $\eta_i$ is defined by
\begin{equation}
    \eta_i = \frac{1}{2} \ln \left( \frac{|\vb{p}_i| + p_{\parallel, i}}{|\vb{p}_i| - p_{\parallel, i}} \right).
\end{equation}
\par\noindent
Given that we are interested in the interactions driving the showers, we focus the following deduction on particles with $p_{\parallel} > 0$, in the laboratory frame. If $p_{\perp, i}$ denotes the transverse momentum of particle $i$, then
\begin{equation}
    \eta_i \simeq \ln \left( \frac{2 E_0}{p_{\perp, i}} \right) + \ln x_i,
\end{equation}
\par\noindent
where we took the ultra-relativistic and small emission angle limits $E_i \simeq p_i \simeq p_{\parallel, i}$. It follows
\begin{equation}
\begin{aligned}
   \zetahad & \simeq - \sum_{i = 1}^{\multhad} x_i \eta_i + \sum_{i = 1}^{\multhad} x_i \ln \left(\frac{2 E_0}{p_{\perp, i}} \right) \\
   \zetaem & \simeq - \sum_{j = 1}^{\multem} x_j \eta_j + \sum_{j = 1}^{\multem} x_j \ln \left(\frac{2 E_0}{p_{\perp, j}}\right)
\end{aligned}
.
\end{equation}
\par
Neglecting fluctuations of $p_{\perp, i}$ when compared to fluctuations of the energy, in the lab. frame, we let $p_{\perp, i} \sim Q$, where $Q$ is the transverse momentum scale of the proton-beam, yields
\begin{equation} \label{eq:zeta_eta_relation}
\begin{aligned}
    \zetahad & \simeq - \sum_{i = 1}^{\multhad} x_i \eta_i + \alphahad \eta_{\text{beam}} \\
    \zetaem & \simeq - \sum_{j = 1}^{\multem} x_j \eta_j + \alphaem \eta_{\text{beam}}
\end{aligned}
,
\end{equation}
\par\noindent
where $\eta_{\text{beam}} = \ln \left(2E_0 /Q\right)$ denotes the proton beam rapidity. Defining the notation $\etahad = - \sum_i x_i \eta_i$, and similarly for the electromagnetic sector yields
\begin{equation}
\begin{aligned}
    \zetahad & \simeq - \etahad + \alphahad \eta_{\text{beam}} \\
    \zetaem & \simeq - \etaem + \alphaem \eta_{\text{beam}}
\end{aligned}
,
\end{equation}
\par
\begin{figure}[h!]
    \centering
    \includegraphics[width=\linewidth]{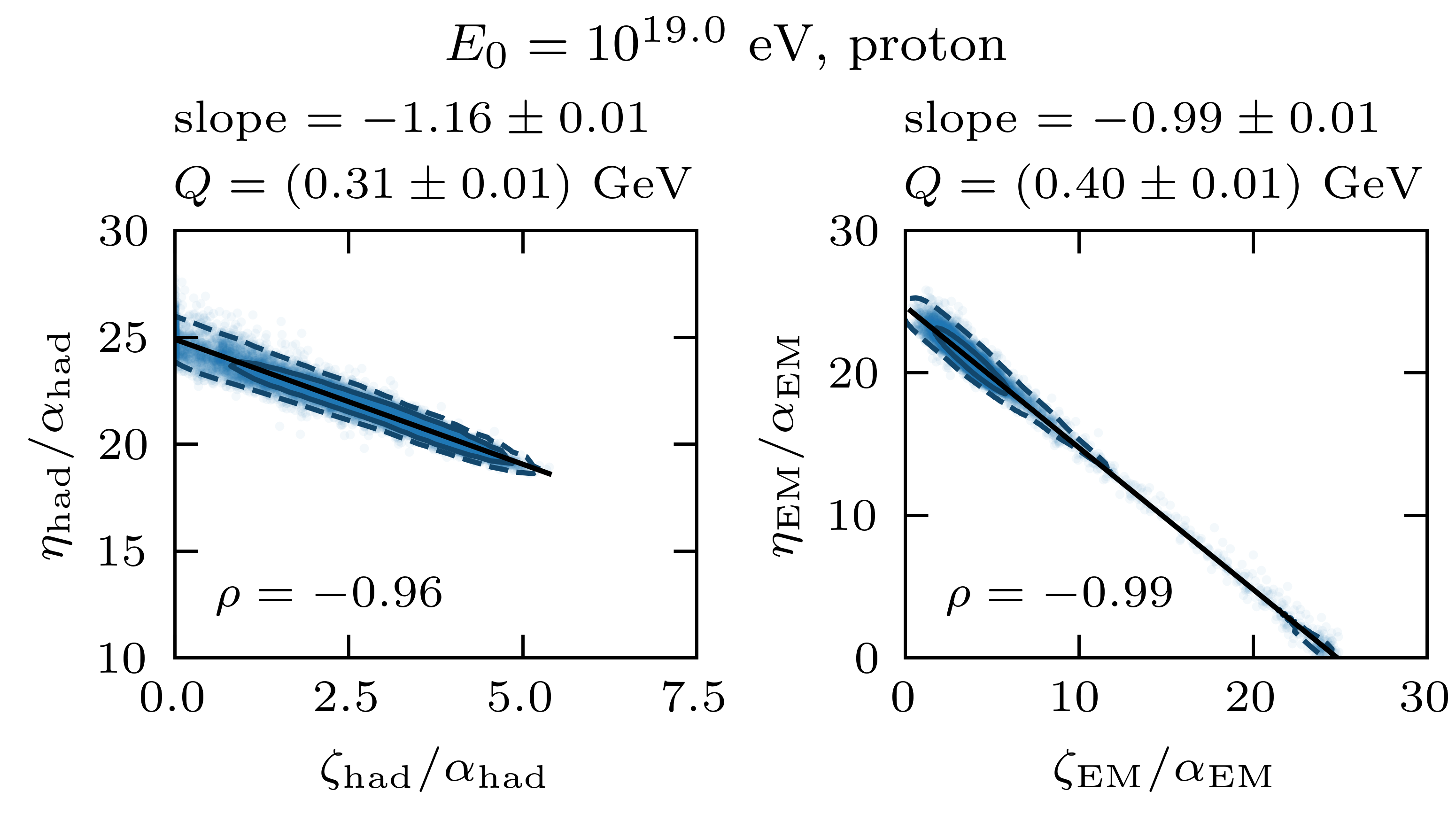}
    \caption{Left panel: joint distribution of $\etahad / \alphahad$ and $\zetahad / \alphahad$. Right panel: joint distribution of $\etaem / \alphaem$ and $\zetaem / \alphaem$. Both panels display the contours containing 68 and $95 \%$ of the events, as dark blue solid and dashed lines respectively, along with a linear regression curve drawn as a solid black line. The values of the slope, $m$, and transverse momentum scale, $Q$, were extracted from the linear regression. The main text details the simulations used to produce the figure.}
    \label{fig:zeta_eta_joint_dist_epos}
\end{figure}

The validity of these approximations is clear from Figure~\ref{fig:zeta_eta_joint_dist_epos}.
\par
We have verified that the Pearson correlation coefficients, $\rho < -0.95$, between $\zetahad / \alphahad$ and $\etahad / \alphahad$ are independent of the hadronic interaction model. The same holds for the electromagnetic component. Moreover, the extracted momentum scales from the linear regression $Q \sim 150 - 250\,$MeV are of the order of magnitude of $p_\perp$~\cite{2011_Engel_HadIntRev, 2023_Cazon_MuonUniversality}. The slope of the linear relation between $\zetahad$ and $\etahad$ is incompatible with one likely due to the mixture of different hadronically interacting particles.

\section{New multiparticle production variables in accelerator experiments} \label{sec:new_variables_accelerators}
Since the quantities $\zetahad$, $\zetaem$ and $\alphahad$ are constructed directly from the energy spectra of secondaries of hadronic interactions, their values can be readily computed in accelerator experiments, in the kinematic phase-space allowed by the particular detector. In particular, the proton-oxygen runs in Run~3 of the LHC~\cite{2019_LHC_pOrun} could provide important information to constrain these variables in proton-air interactions in the atmosphere.

\par
\begin{figure}[h!]
    \centering
    \includegraphics[width= \linewidth]{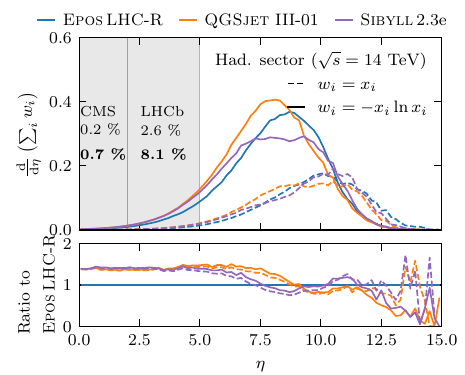}
    \caption{Upper panel: Contribution of each particle to the values of $\alphahad$ (solid lines) and $\zetahad$ (dotted lines) as a function of the particle's pseudo-rapidity for different hadronic interaction models. Lower panel: ratio to the energy and $\zeta$ flows predicted by \eposr{}. Proton-air interactions were simulated with $E_0 = 10^{17}\,$eV corresponding to $\sqrt{s} = 14\,$TeV. The shaded grey bands represent the pseudo-rapidities covered by CMS and LHCb. The percentages in roman and bold font type refer to the fraction of the energy and $\zeta$-flows, respectively, covered by each detectors.}
    \label{fig:zeta_had_flow_17}
\end{figure}

The forward region of the kinematic phase-space contributes the most for the values of $\zetahad$ and $\alphahad$, on an event-by-event level, as seen in Figure~\ref{fig:zeta_had_flow_17}. This figure was produced from proton-air interactions at $E_0 = 10^{17}\,$eV, corresponding to the nucleon-nucleon centre-of-mass energy $\sqrt{s} = 14\,$TeV. The corresponding results for interactions at $E_0 = 10^{18.7}\,$eV $\implies \sqrt{s} = 97\,$TeV, the approximate center-of-mass energy proposed for the Future Circular Collider~(FCC)~\cite{2019_Abada_FCC} are discussed in Appendix~\ref{apx:zeta_flow}. For reference, the pseudo-rapidity regions covered by the particle detectors CMS, $\eta < 2$~\cite{2008_CMS_design}, and LHCb, $2 < \eta < 5$~\cite{2008_LHCb_designReport}, are represented by the light grey vertical bands. The contribution of each hadronically interacting particle to $\alphahad$, in each pseudo-rapidity bin, $\dd{\eta}$, is the differential energy flow of the hadronic component of the shower. By analogy, the contribution to $\zetahad$ per pseudorapidity bin will be referred to as $\zeta$-flow. In Figure~\ref{fig:zeta_had_flow_17}, the fraction of the integrated energy flow in the pseudo-rapidity regions covered by each detector is represented by the numbers in Roman type within the shaded bands. The integrated $\zeta$-flow is represented by the numbers in bold type within the shaded bands.
\par
The energy flow to the hadronic sector peaks at higher rapidities than the $\zeta$ flow. This is expected since the extreme rapidity regions are populated by particles produced in \textit{quasi}-elastic and diffractive interactions, where $\zetahad \to 0$ and $\alphahad \to 1$. Both peaks lie outside the region covered by current detectors. However, they lie inside the rapidity region covered by the detectors of Forward Physics Facility (FPF)~\cite{2023_FPF_designReport, 2021_Kling_FPFneutrinos}: $\eta > 7.2$. The FASER and FLARE experiments of this facility will be able to probe the energy spectrum of charged pions and kaons via their decays into neutrinos. Since these charged particles contribute, on average, to $\sim 60 \%$ and $\sim 20 \%$ of the values of $\zetahad$, measuring their spectra would constrain $\zetahad$ in the kinematic region where particles most contribute to $\zetahad$. Above $\eta = 5$, the shapes of the energy and $\zeta$-flows are highly dependent on the hadronic interaction model, especially at the highest rapidities, where accelerator data do not constrain the models.
\par
\begin{figure}[h!]
    \centering
    \includegraphics[width= \linewidth]{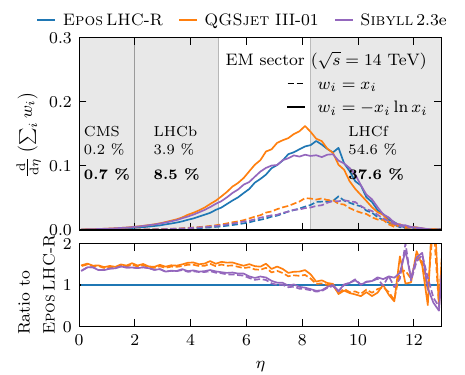}
    \caption{Upper panel: Contribution of each particle to the values of  $1 - \alphahad$ (solid lines) and $\zetaem$ (dotted lines) as a function of the particle's pseudo-rapidity for different hadronic interaction models. Lower panel: ratio to the energy and $\zeta$ flows predicted by \eposr{}. Proton-air interactions were simulated with $E_0 = 10^{17}\,$eV corresponding to $\sqrt{s} = 14\,$TeV. The shaded grey bands represent the pseudo-rapidities covered by CMS,  LHCb and LHCf. The roman and bold type percentages refer to the energy and $\zeta$-flows, respectively, covered by the detectors.}
    \label{fig:zeta_em_flow_17}
\end{figure}

The corresponding $\zeta$-flow for the electromagnetic component is shown in Figure~\ref{fig:zeta_em_flow_17}, for proton-air interactions at $\sqrt{s} = 14\,$TeV, along with the ratio to the $\zeta$-flow predicted by \eposr{}. The corresponding flows for proton-air interactions at $\sqrt{s} = 97\,$TeV are discussed in Appendix \ref{apx:zeta_flow}. The figure displays an additional band showing the pseudo-rapidities covered by the LHCf detector~\cite{2008_LHCf_design}, able to measure the spectra of neutral pions and neutrons in the pseudo-rapidity region $8.3 < \eta < 13$~\cite{2016_LHCf_forward_npions, 2020_LHCf_forwardneutrons}.
\par
Most features of the $\zeta$-flow for the hadronic component are also present in the $\zeta$-flow for the EM component. About $40 \%$ of the integrated $\zeta$-flow can be measured at the LHCf, which additionally covers $~55\%$ of the primary energy flowing into forward neutral pions. In this far-forward pseudo-rapidity range, the $\zeta$-flow is model-dependent in the lower rapidity range covered by LHCf. However, better discrimination would be achieved for higher centre-of-mass energies in human-made accelerators, as discussed in Appendix~\ref{apx:zeta_flow} or using UHECRs measurements. Therefore, both accelerator and cosmic-ray experiments offer complementary relevant information.
\par
\begin{figure*}[th!]
    \centering
    \includegraphics[width=\linewidth]{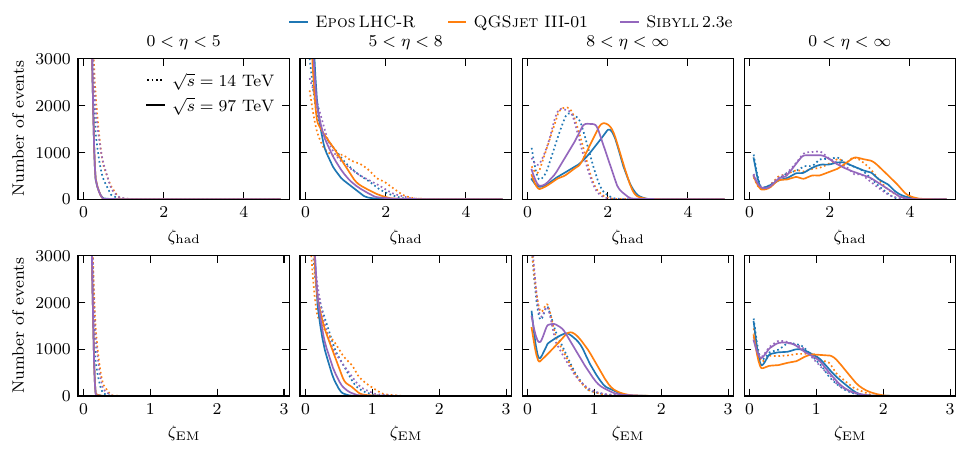}
    \caption{Distributions of $\zetahad$ (upper panels) and $\zetaem$ (lower panels) for different pseudo-rapidity regions. Distributions obtained at $\sqrt{s} = \SI{14}{\TeV}$ ($E_0 = 10^{17}\,$eV) are depicted as dotted lines, while those obtained at $\sqrt{s} = \SI{97}{\TeV}$ ($E_0 = 10^{18.7}\,$eV) are depicted as solid lines. The rightmost panels correspond to the rapidity range covered in p-air interactions in Extensive Air Showers.}
    \label{fig:zetahad_eta_bins_both_energies}
\end{figure*}

The distributions of $\zetahad$ and $\zetaem$ can be directly built from the measurement of the energy spectra of secondaries within each pseudo-rapidity bin. Figure ~\ref{fig:zetahad_eta_bins_both_energies} shows these distributions for two nucleon-nucleon center-of-mess energies $\sqrt{s} = \SI{14}{\TeV}$ and $\sqrt{s} = \SI{97}{\TeV}$, and four pseudo-rapidity bins: $0 < \eta < 5$ (combined coverage of CMS and LHCb), $5 < \eta < 8$ (not covered by any of the detectors), $8 < \eta < \infty$ (covered by LHCf and potentially FPF), and the entire $\eta$-range as covered in extensive air showers.

\par
The differences between hadronic interaction models become more pronounced with increasing pseudo-rapidity and interaction energy, highlighting the potential for future high-energy experiments such as the FCC to measure the particle energy spectra in the kinematic regimes most relevant for EAS. The LHCf measurements of the neutral pion energy spectrum in the forward region greatly constrain the shape of the distribution of $\zetaem$ at $\sqrt{s} = \SI{14}{\TeV}$. However, discrepancies between the models persist at the highest energies, where no direct data is available. In the far-forward region ($\eta > 8$), at lower energies, the distributions of $\zetahad$ obtained with \qIII{} and \sibe{} are very similar, whereas that obtained with \eposr{} differs significantly. This suggests that the introduction of hadron re-scattering, which alters the connection between mid and forward rapidity, impacts the shape of $\zetahad$. Notably, the largest discrepancies between models appear in the rapidity range accessible only through extensive air shower observations, underscoring the importance of cosmic-ray studies in constraining hadronic interaction models.
\par
The potential explored in this paper adds to the important steps in accessing the information about hadronic interactions in EAS, in a data-driven way, presented in ~\cite{2012_Auger_xsection, 2023_Olena_xsection_comp, 2018_Cazon_alpha, 2021_Cazon_lambdamu, 2023_Goos_HadInt_xmaxnmu, 2024_Martins_lambdamu_xmax}. Moreover, the potential for accelerator measurements in combination with constraints on hadron production via the distribution of $\xmax$ can significantly mitigate the tension between EAS simulations and measurements.

\section{Summary and outlook} \label{sec:conclusions}

In this work, we introduced the variable, $\xi$, which governs the majority of fluctuations in $\Delta X_{\max}$ during ultra-high-energy proton-air interactions. This variable is constructed as a linear combination of more fundamental quantities, $\zetahad$, $\zetaem$, and $\alphaem$, derived directly from the energy spectra of secondary particles produced in the first p-Air interaction.

\par We demonstrated that the event-by-event value of $\xi$ provides a robust estimation of $\Delta X_{\max}$ with minimal dependence on the hadronic interaction model. Notably, fluctuations in $\xi$ account for approximately $50\%$ of the total $\Delta X_{\max}$ fluctuations—remarkably close to the upper limit of $65\%$ that would be obtained if all first-interaction information were retained. This confirms that fluctuations from subsequent shower development play only a secondary role in shaping the $\Delta X_{\max}$ distribution.

\par To quantify the shower evolution beyond the first interaction, we parameterized the residual fluctuations in $\Delta X_{\max}$ using a probabilistic kernel. The consistent change of the kernel due to changes in the spectra of the primary interaction was derived from the available high-energy interaction models.  This provides a universal framework that directly connects hadronic interaction properties with extensive air shower measurements. Under this framework, a prior distribution of $\xi$ allows for the prediction of the corresponding distribution of $\xmax$ with biases in its main moments of less than $\SI{3}{\depth}$. Therefore, a data-driven measurement of the distribution of $\xmax$ could be used to constrain, with minimal dependence on the hadronic interaction model, the energy spectra of secondaries in ultra-high-energy hadronic interactions.


\par We analyzed how $\zetahad$, $\zetaem$, and $\alphahad$ can be extracted from different rapidity regions accessible to current particle detectors. The dependence on the hadronic interaction model increases with both the interaction’s center-of-mass energy and the rapidity coverage of the measurements. While the energy flow contribution from different rapidity regions is highly model-dependent and can be directly evaluated using existing detector data, we found that the event-by-event distributions of $\zetahad$ and $\zetaem$ are only distinguishable in the far-forward region of the kinematic phase space. These variables are most distinguishable in the phase space relevant for EAS development.

\par In conclusion, extensive air showers represent the best available tool for constraining multiparticle production in hadronic interactions. This can be achieved through the explicit probabilistic link between the fundamental variables $\{\zetahad, \zetaem, \alphahad \}$ and the observable $X_{\max}$.
\par
Applying the framework developed in this work to data is not trivial, as the mass composition of the ultra-high-energy cosmic ray flux, as interpreted by the hadronic interaction models, is compatible with a mixture of light and heavy nuclei~\cite{2016_Auger_XmaxS38corr}. One possible strategy is to exploit correlations between shower observables, such as $\xmax$ and the number of muons, and restrict the analysis to regions of their joint distribution dominated by proton primaries. For this subset of events, typically characterised by deep $\xmax$ values and low muon content, the framework can be applied without modification. Alternatively, the correspondence between $\xi$ and $\xmax$ could be extended to heavier nuclei. This would, in principle, require a precise probabilistic treatment of nuclear fragmentation and an evaluation of the sensitivity of the $\xmax$ distribution to collective hadronisation effects in the initial interactions. Nevertheless, this work constitutes an initial step in a phenomenological exploration of the production properties of the primary interaction via the full distributions of EAS observables.
\par

\section*{Acknowledgments}
The authors thank Silvia Mollerach and Gonzalo Parente for carefully reading this manuscript. We extended our gratitude to the Auger-IGFAE, Auger-LIP, and the Pierre Auger Collaboration members for their valuable insights throughout the different stages of this work. The authors thank Ministerio de Ciencia e Innovaci\'on/Agencia Estatal de Investigaci\'on
(PID2022-140510NB-I00 and RYC2019-027017-I), Xunta de Galicia (CIGUS Network of Research Centers,
Consolidaci\'on 2021 GRC GI-2033, ED431C-2021/22 and ED431F-2022/15),
and the European Union (ERDF). This work has been partially funded by Fundação para a Ciência e Tecnologia, Portugal, under project \url{https://doi.org/10.54499/2024.06879.CERN}. MAM acknowledges that the project that gave rise to these results received the support of a fellowship from ``la Caixa” Foundation (ID 100010434). The fellowship code is LCF/BQ/DI21/11860033.
F.R. has received funding from the European Union’s Horizon 2020 research and innovation programme under the Marie Skłodowska-Curie grant agreement No. 101065027.

\bibliography{references}

\appendix

\section{Impact of diffractive primary interactions in the tail of \boldmath{$\dxmax$}} \label{apx:leading_fluctuations}

\begin{figure}[h!]
    \centering
    \includegraphics[width=\columnwidth]{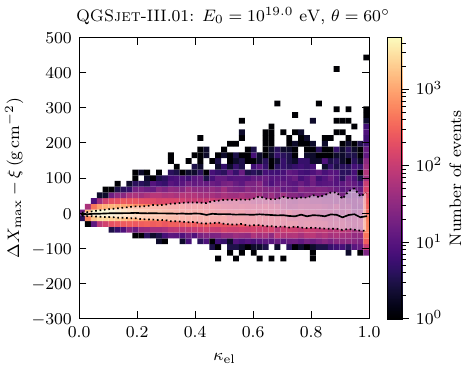}
    \caption{Distribution of the residuals $\dxmax - \xi$ as a function of the elasticity of the first $p$-air interaction $\elasticity$. The black line represents the bias $\expval{R_X}$ as a function of $\elasticity$. The dotted lines correspond to the asymmetrical resolution. This figure was produced with the library of proton-induced \conex{} described in Section~\ref{sec:xi_model_performance}, using \qIII{}.}
    \label{fig:simpleXi_vs_dXmax_elasticity_residuals}
\end{figure}

In the case of highly elastic primary interactions, there is little information in the first $p$-air interaction to explain the fluctuations in shower observables. Furthermore, the fluctuations of the interaction depths of hadronically interacting particles are less suppressed due to the low multiplicity of secondaries produced in such interactions~\cite{2024_Martins_lambdamu_xmax}. This leads to increased variability in the residuals $R_X = \dxmax - \xi$ with increasing elasticity of the first interaction, $\elasticity$, as shown in Figure~\ref{fig:simpleXi_vs_dXmax_elasticity_residuals}. The dotted black lines correspond to the asymmetrical resolution in the prediction of $\dxmax$ in a given bin of $\elasticity$, while the solid black line represents the bias.
\par
\par
\begin{figure}[h!]
    \centering
    \includegraphics[width=\columnwidth]{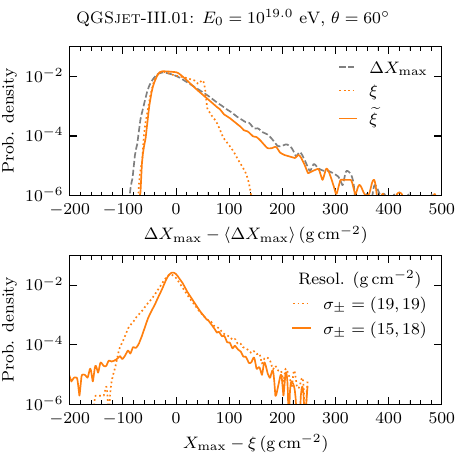}
    \caption{Top panel: distributions of $\xi$ (dotted blue), $\widetilde{\xi}$ (solid blue) and $\dxmax$ (dashed grey) for $10^4$ \conex{} proton-induced showers with $E_0 = 10^{19}$ eV and $\theta = 67^\circ$, using \qIII{}. Bottom panel: distributions of residuals of $\xi$ (dotted line) and $\widetilde{\xi}$ (solid line), defined in Equations~\eqref{eq:xi_model_simple} and \eqref{eq:xi_model_minimal_extension}, relative to $\dxmax$. The respective resolutions can be read in the legend.}
    \label{fig:minimal_extension_Xi_residuals_qIII}
\end{figure}

Letting $\lambda_\ell$ denote the actual depth travelled by the leading of the first interaction before interacting again, we add a correction term to $\xi$ (see Equation~\eqref{eq:xi_model_simple}) to account for the exponential fluctuations of $\lambda_\ell$. We thus obtain
\begin{equation} \label{eq:xi_model_minimal_extension}
    \widetilde{\xi} = \xi + \elasticity \left( \lambda_\ell - \lambda(\kappa_{\text{el}})\right),
\end{equation}
\par\noindent
where $\lambda(\elasticity) = \lambda_r\left[\lambda_0 - \delta \ln \left(\elasticity E_0 / E_{\text{ref}} \right)\right]$. For a subset of $10^5$ \conex{} simulations, the distributions of $\dxmax$, $\xi$ and $\widetilde{\xi}$ are shown in the top panel of Figure~\ref{fig:minimal_extension_Xi_residuals_qIII}, while the bottom panel shows the distributions of residuals for $\xi$ (dashed light grey) and $\widetilde{\xi}$ (solid blue). The free parameters in Equation~\ref{eq:xi_model_minimal_extension} were tuned as described in Section~\ref{subsec:xi_model_tuning}, and averaged over the three hadronic interaction models.
\par
\par
The distribution of $\widetilde{\xi}$ has a more pronounced exponential high tail, offering a better description of the full distribution of $\dxmax$ when compared to the distribution of $\xi$. Additionally, the peak caused by diffractive events present in the distribution $\xi$ is largely absent in the distribution of $\widetilde{\xi}$, as it includes the fluctuations of the interaction length of the leading proton. Moreover, the distribution of residuals of $\widetilde{\xi}$ is more symmetrical, narrower, and has a less pronounced high tail, leading to better resolutions of $\sigma_{\pm} = _{15}^{18}\unit{\depth}$. The correlation coefficient between $\widetilde{\xi}$ and $\dxmax$ was determined to be $0.82$, a clear improvement relative to $\xi$. We infer that the exponential tail in $\dxmax$ comes mostly from the exponential fluctuations of the interaction depth of the leading particle of the primary interaction, with this effect being increasingly more relevant with the elasticity of the primary interaction. Therefore, the loss of correlation between $\xi$ and $\dxmax$ for highly elastic primary interactions is not dominated by particle production in later shower generations, but rather the fluctuations of the interaction depth of the leading of the first generation.
\section{Energy dependence of parameters in the definition of \boldmath{$\xi$}} \label{apx:xi_params_energy_func}
Table~\ref{tab:xi_params_energy_func} shows the values of the parameters used in Equation~\eqref{eq:omega_energy_func}, as a function of the primary energy $E_0$.
\begin{table}[ht!]
    \centering
    \caption{Primary-energy dependence of the optimal parameters used in Equation~\eqref{eq:omega_energy_func}.}
    \label{tab:xi_params_energy_func}
    \begin{tabular}{c|c|c}
    \hline\hline
    $\log_{10}(E_0 / \unit{\eV})$ & $p_1$ & $p_0$ \\ \hline
    $17.0$ & \num{-0.324(3)} & \num{1.57(1)} \\
    $17.5$ & \num{-0.316(4)} & \num{1.64(1)} \\
    $18.0$ & \num{-0.222(3)}  & \num{1.34(1)} \\
    $18.5$ & \num{-0.153(3)}  & \num{1.12(1)} \\
    $19.0$ & \num{-0.140(4)}  & \num{1.09(1)}  \\
    $19.5$ & \num{-0.122(4)} & \num{1.04(1)}  \\
    $20.0$ & \num{-0.090(4)}  & \num{0.93(1)}  \\ \hline \hline
    \end{tabular}
\end{table}
\section{Energy and $\zeta$-flows at $E_0 = 10^{18.7}$ eV}
\begin{figure}[h!]
    \centering
    \includegraphics[width= \linewidth]{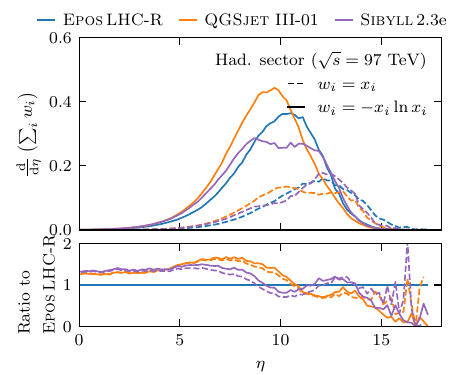}
    \caption{Upper panel: Contribution of each particle to the values of $\alphahad$ (solid lines) and $\zetahad$ (dotted lines) as a function of the particle's pseudo-rapidity, that is, the energy and $\zeta$-flows respectively, for different hadronic interaction models. Lower panel: ratio to the energy and $\zeta$ flows predicted by \eposr{}. Proton-air interactions were simulated with $E_0 = 10^{18.7}$ eV corresponding to $\sqrt{s} = 97$ TeV}
    \label{fig:zeta_had_flow_19}
\end{figure}

\begin{figure}[h!]
    \centering
    \includegraphics[width= \linewidth]{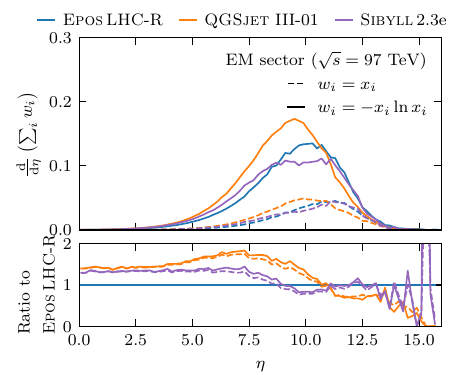}
    \caption{Upper panel: Contribution of each particle of the electromagnetic sector to the values of $\zetaem$ as a function of the particle's pseudo-rapidity, that is the $\zeta$-flow, for different hadronic interaction models. Lower panel: ratio to the energy and $\zeta$ flows predicted by \eposr{}. Proton-air interactions were simulated with $E_0 = 10^{18.7}$ eV corresponding to $\sqrt{s} = 97$ TeV.}
    \label{fig:zeta_em_flow_19}
\end{figure}

The energy and $\zeta$-flows into the hadronic sector of proton-air interactions at a centre-of-mass energy of $\sqrt{s} = \SI{97}{\tera\electronvolt}$, the nucleon-nucleon centre-of-mass energy projected for the FCC~\cite{2019_Abada_FCC}, is shown in Figure~\ref{fig:zeta_had_flow_19}. The definition of $\zeta$-flow can be found in Section~\ref{sec:new_variables_accelerators}. The corresponding flow into the electromagnetic sector can be found in Figure~\ref{fig:zeta_em_flow_19}.


\label{apx:zeta_flow}
\par

\end{document}